\documentclass[]{article}
\usepackage{cite}
\usepackage{amsmath,amssymb,amsfonts}
\usepackage{algorithmic}
\usepackage{graphicx}
\usepackage{textcomp}

\usepackage{subfig}
\DeclareGraphicsExtensions{.pdf,.jpeg,.png}
\usepackage{url}
\usepackage{multirow}
\usepackage{color}
\usepackage{array}
\usepackage{subfig}
\usepackage{bm}

\usepackage{booktabs}
\usepackage{threeparttable}

\begin{document}
\title{Multi-Modality Cardiac Image Analysis with Deep Learning}
\author{Lei Li$^\#$ , Fuping Wu$^\#$, Sihang Wang$^\#$, Xiahai Zhuang$^\#$
\thanks{\# All authors contribute equally to this article.}
\thanks{Lei Li is with Department of Engineering Science, University of Oxford, Oxford, UK.
  Fuping Wu, 
  Sihan Wang and
  Xiahai Zhuang are with School of Data Science, Fudan University, Shanghai, China.
  Xiahai Zhuang is the corresponding author, zxh@fudan.edu.cn.}}

\date{}
\maketitle

% \begin{frontmatter}

% \chapter{Multi-Modality Cardiac Image Analysis with Deep Learning}\label{chap1}
% %\subchapter{DL-based MM Cardiac Image Computing}
% \begin{center}
% \subchapter{\small{Lei Li$^\#$, Fuping Wu$^\#$, Sihang Wang$^\#$, Xiahai Zhuang$^\#$}\footnote{$\#$ All authors contribute to this chapter equally.}}
% \end{center}

\begin{abstract}
Accurate cardiac computing, analysis and modeling from multi-modality images are important for the diagnosis and treatment of cardiac disease.
Late gadolinium enhancement magnetic resonance imaging (LGE MRI) is a promising technique to visualize and quantify myocardial infarction (MI) and atrial scars.
Automating quantification of MI and atrial scars can be challenging due to the low image quality and complex enhancement patterns of LGE MRI.
Moreover, compared with the other sequences LGE MRIs with gold standard labels are particularly limited, which represents another obstacle for developing novel algorithms for automatic segmentation and quantification of LGE MRIs.
This chapter aims to summarize the state-of-the-art and our recent advanced contributions on deep learning based multi-modality cardiac image analysis.
Firstly, we introduce two benchmark works for multi-sequence cardiac MRI based myocardial and pathology segmentation.
Secondly, two novel frameworks for left atrial scar segmentation and quantification from LGE MRI were presented.
Thirdly, we present three unsupervised domain adaptation techniques for cross-modality cardiac image segmentation.

\par
\emph{Multi-Modality, Cardiac Image, Deep Learning, Domain Adaptation
}\rm
\end{abstract}

% \begin{keywords}
% \kwd{Multi-Modality}
% \kwd{Cardiac Image}
% \kwd{Deep Learning}
% \kwd{Domain Adaptation}
% \end{keywords}

% \end{frontmatter}

% \section{Chapter points}\label{points}
% \begin{itemize}
%     \item[1.] Combining multi-modality images is important for cardiac analysis and can enlarge the size of training data for deep learning based algorithms.   
%     \item[2.] The complementary information from multi-sequence CMR images can be helpful for both cardiac structure and pathology segmentation. 
%     \item[3.] The left atrial scar quantification from LGE MRI based on surface projection is promising and can be potentially useful in the diagnosis and prognosis of atrial fibrillation. 
%     \item[4.] There domain adaptation algorithms were proposed, including feature disentanglement, explicit discrepancy metrics "CF-distance", and VarDA that drives two domains towards a common parameterized distribution.
% \end{itemize}

\section{Introduction}\label{intro}
%multi-modality cardiac image
Multi-modality cardiac images are widely utilized to assist the diagnosis and treatment management of patients.
Cardiac computed tomography (CT) is generally fast, low cost and with high quality for cardiac imaging.
Magnetic resonance imaging (MRI) can provide important anatomical and function information of the heart, and different sequences of MRI can further capture different information.
For example, the LGE MRI sequence has been widely used to visualize myocardial infarction (MI) as well as left atrial (LA) fibrosis and scars;
the balanced-steady state free procession (bSSFP) cine sequence can present clear cardiac boundaries and captures cardiac motions in different phases;
the T2-weighted MRI sequence can display the acute injury and ischemic regions.
Combining multi-modality/sequence cardiac images is a promising research direction in the literature, such as multi-modality registration \cite{conf/IPMI/collignon1995}, multi-modality/sequence fusion \cite{journal/TMI/menze2014,journal/Array/zhou2019}, and domain adaptation \cite{journal/MedIA/tajbakhsh2020}.

In this chapter, we will present three topics related to the multi-modality cardiac image processing.
Section \ref{MSCMR}  presents two challenges for multi-sequence cardiac MRI based myocardial and pathology segmentation.
Section \ref{LA}  presents two novel frameworks, namely LearnGC and AtrialJSQnet, for left atrial scar segmentation and quantification from LGE MRI (with the assist of an additional non-enhanced MRI).
Section \ref{DA}  introduces three unsupervised domain adaptation algorithms for cross-modality cardiac image segmentation.

%application 1: myocardial pathology segmentation
%application 2: LA scar quantification
%method: domain adaptation

\section{Multi-sequence cardiac MRI based myocardial and pathology segmentation}\label{MSCMR}
\subsection{Introduction}
% According to WHO, Cardiovascular disease(CVDs) are the leading cause of death globally, causing an estimated 17.9 million death each year, among which 
Cardiovascular disease is one of the leading causes of global death, among which MI is the most acute and deadly one \cite{thygesen2007universal}. 
Early diagnosis as well as prompt treatment are the key to prevent the poor prognosis.
Among the imaging modalities available in clinical routines for MI, cardiac magnetic resonance (CMR) imaging becomes the gold-standard technique. 
The functional cardiac indexes, i.e., left ventricular (LV), ejection fraction (EF), LV volumes, myocardium mass as well as precise location of lesion could be obtained from the analysis of MRI.
However, one MRI sequence could  provide limited clinic information. 
In clinic practice, multi-modality would be simultaneously utilized, since the complementary information from multi sequences can assist the diagnosis and location of lesion. 
Whereas, the misalignment among multi modalities caused by the different scanning orientations, the various sequences characteristic and the low tissue contrast in special modalities (such as T2 and LGE CMR) make it challenging to handle multi-modality tasks.

\begin{figure*}[t]\center
    \includegraphics[width=\textwidth]{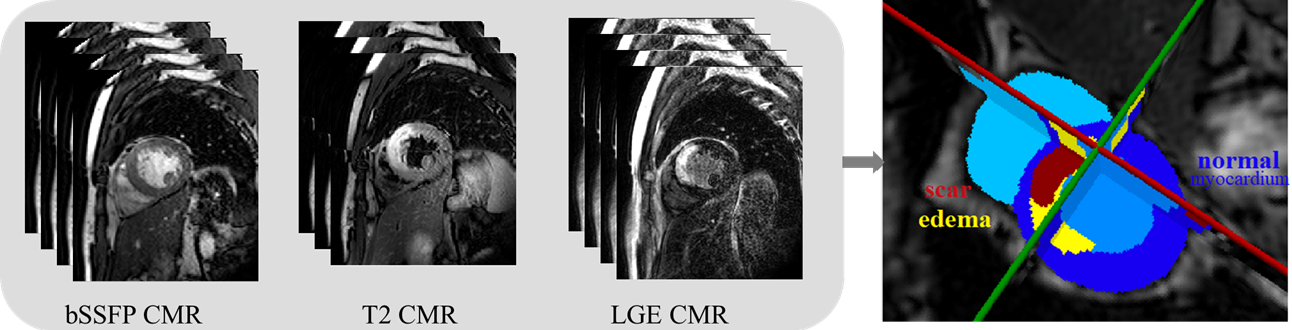}
  \caption{Visualization of myocardial pathology segmentation combing multi-sequence CMR images acquired from the same patient. Image adopted from the website of MyoPS 2020 challenge: http://www.sdspeople.fudan.edu.cn/zhuangxiahai/0/myops20/.}
\label{fig:multi-sequences}
\end{figure*}

%In recent years, we have been studying the segmentation and registration task using multi-modality CMR data.
% First,  the multivariate mixture model (MvMM) were proposed to adjust the misalignment among multi-modalities caused by the different scanning orientations, and achieve anatomical structure segmentation \cite{zhuang2018multivariate}. 
%To keep the chapter concise, we will solely elaborate on segmentation methods, taking two segmentation challenge held by us as example. 
To boost the study in this field, in recent years, we have organized two challenges, providing a fair and unified platform and open benchmark dataset for researches around the world.
We first held MS-CMRSeg challenge event for LGE CMR image segmentation with multi-sequence CMR available, segmenting the anatomical structure (i.e., left ventricle, right ventricle and myocardium )  . 
Fig. \ref{fig:multi-sequences} provides an example of multi-sequence images from one subject, including bSSFP, T2 and LGE CMR.
One can see that there exist misalignment, vast image intensity distribution gap and various pathology region shape among them.
These differences are the main obstacles for image analysis. 
As an extension, we then organized myocardial pathology segmentation (MyoPS) challenge event for the myocardial pathology segmentation, including scar and edema.
In the following, we summarize the top-performed methods in the two challenges.

\subsection{Methodology summary for challenge events}
% \subsection{Method}
% \subsubsection{Multicomponent GMM and coupled level set (2017 TBME)}
% \subsubsection{LGE MRI segmentation incorporating shape and spatial priors (2019 MICCAI)}

% \Leicolor{}{lei: maybe it would be better to cite the paper of each team, instead of just using team *** to represent their method.  \\wsh: improved~---lei: not just table but also the main body.}

\subsubsection{MS-CMRSeg Challenge: Cardiac Segmentation on Late Gadolinium Enhancement MRI} \label{challenge:MSCMR2019}
We organized multi-sequence cardiac MR segmentation (MS-CMRSeg) challenge, in conjunction with MICCAI 2019.
The challenge mainly focused on the anatomical structure (ventricles and myocardium) segmentation of LGE sequence with complementary information from the other two sequences (bSSFP and T2 sequence). 
Sixty-five teams from all over the world participated in the challenge, twenty three of which submitted results for evaluation before the submission deadline. 
To keep the chapter concise, only the top nine methods will be discussed. 
These nine methods could roughly be divided into two categories.
% \textcolor{red}{(wfp: interpretation for what?)} 
The first one is training without LGE annotation, namely an unsupervised domain adaptation (UDA) problem, and three teams focused on this task;
the other six teams trained their models with a few LGE ground-truth, considered as supervised methods. 
% \textcolor{red}{(wfp: The corresponding methods could be categorized into either UDA algorithm or supervised approach?)} 

% \textcolor{red}{(wfp:but you also described supervised methods in next paragraph! )} 
For UDA methods, two teams, i.e., ICL \cite{conf/STACOM/chen2019} and INRIA \cite{mscmr/nvidia}, adopted image translation scheme via the style-transfer. 
ICL leveraged multi-modal unsupervised image-to-image translation network \cite{huang2018multimodal} to implement style transfer. 
In their framework, the encoder extracts style information and structure information from image separately, while decoders reconstruct the image using the extracted features. 
%Intuitively, combining source style feature and target structure feature could generate image with style of source domain and content similar to the target image.
INRIA adopts conventional style transformation, i.e., histogram matching towards LGE, and also employs some data augmentation schemes including adaptive histogram equalization and intensity inversion.
As for the network structure, ICL adopts two stage cascaded network to extract "coarse to fine" features, while INRIA utilizes a dual U-Net \cite{jia2018automatically}. 
Instead of using image translation, XMU \cite{MSCMR/XIAN} performs feature alignment to achieve UDA optimization.
The alignment is achieved via a discriminator, which can minimize the distribution discrepancy between the two domains in both feature and output spaces.

%Different from the aforementioned teams, the majority of the participants focused on supervised task. 
% \textcolor{red}{(wfp:Table 1.1 and 1.2 are not explained?! As presented in Table 1.1, the other top six teams adopted supervised learning strategies. )} 
Among the supervised-learning based methods, UB \cite{MSCMR/UB} and HIT \cite{mscmr/HIT} mainly adopted several date augmentation strategies to avoid over-fitting, while the other teams try to design effective model structures for LGE MRI segmentation. 
For data augmentation, UB leverages CycleGAN strategy \cite{zhu2017unpaired} to convert bSSFP images into LGE-like ones and utilizes region rotation of scars, while HIT uses histogram matching for augmentation.
% \textcolor{red}{(wfp: why do you describe this team for pre-processing?  )}. 
For model structure, only NVIDIA \cite{mscmr/nvidia} leverages conventional multi-atlas segmentation (MAS) framework, while the other teams, i.e., FAU \cite{mscmr/FAU}, SUST \cite{MSCMR/SUST} and SCU \cite{conf/STACOM/wang2020} employs deep neural networks. 
SCU \cite{conf/STACOM/wang2020} redesigns the baseline U-Net with the squeeze-and-excitation residual module \cite{wang2019sk} and the selective kernel module \cite{hu2018squeeze} to recalibrate channel-wise feature responses and adjust the size of its receptive field, respectively. 
FAU \cite{mscmr/FAU} uses transfer-learning to facilitate modeling ability with complementary modality information for bSSFP and T2. 
SUST utilizes generator and discriminator to generate segmentation masks from LGE CMR image. 
% Since there are apparent discrepancies in the distributions of multi-modal images, the leading factor contributing to the results is data pre-processing and augmentation strategy, which could reduce the distribution gap and avoid over-fitting, respectively. \textcolor{red}{(wfp:does the causality hold?  )} 
%  \textcolor{red}{For data augmentation,} 
% style transfer-synthesis image strategy was commonly used. 
% For example,

% As for training strategy, FAU \cite{mscmr/FAU}, while SUST \cite{MSCMR/SUST} proposed an adversarial segmentation framework to drive outputs similar to the ground truth.
% \textcolor{red}{(wfp: why do you describe this team for training strategy? )} 
% In addition, NVIDIA \textcolor{red}{(wfp: reference?)} combined conventional segmentation technique, i.e., multi-atlas label fusion, with deep learning for more robust segmentation. 
% For model architecture, U-Net and its variants are the most prevailing backbone among all the participants. 

% Consonant with prevailing preferred UDA strategies, that is feature invariant and image invariant, the three unsupervised methods can be roughly categorized into two strategy. The essential thought of Feature invariant is that the 
% The three unsupervised methods can be categorized as unsupervised domain adaptation problem. There are two main structure for UDA, According to [cite: DA survey], one

% tb:result:Dice

\subsubsection{MyoPS: Myocardial Pathology Segmentation from Multi-Sequence Cardiac MRI} \label{challenge:MyoPS2020}
We further organized MyoPS challenge as the extension of MS-CMRSeg challenge, in conjunction with MICCAI 2020, to focus on the pathology segmentation instead of the anatomical structure segmentation.
Up to now, MyoPS challenge has received seventy-six requests of registration around world, among which twenty-three teams submitted results for evaluation. 
%Though the challenge is finish, the datasets are still accessible once you register. 
As mentioned before, one of the major challenges of multi-modal image analysis is the misalignment between different modalities. 
To alleviate this, in this challenge we pre-processed the MS-CMR dataset via the MvMM algorithm \cite{journal/PAMI/zhuang2018} for inter-sequence registration.
In this section, we will summarize useful training strategies and trends observed from 15 approaches submitted to MyoPS, in terms of preprocessing, data augmentation, segmentation model architecture and post-processing.

% \textcolor{red}{(wfp:please summarize  preprocessing, data augmentation, segmentation model architecture and post-processing, separately in four paragraphs, each for one topic. and again, what is Table 1.2 used for if you do not cite it in your manuscript?!!!)}

\textbf{Preprocessing.} 
Preprocessing is a crucial technique to mitigate undesirable variations from raw data and reduce modeling complexity. 
Since the region of interest (edema and scar) of CMR is relatively small compared to background, all approaches crops the training images as a preprocessing.
Most of them roughly crops the training images to generate the center of heart. 
For example, USTB \cite{myops/USTB} simply crops the images into the ROI with 256 $\times$ 256 pixels. 
%several team adopted U-Net to localize the position of anatomical structure (i.e,.LV and Myo), then utilized the predictions as masks to crop the original images.
Besides, several teams employs prior segmentation network to automatically localize the position of LV and myocardium for ROI extraction. 
For example, UBA \cite{myops/UBA} leverages U-Net to segment myocardium and crops the smallest bounding box containing the prediction. 
In this way, one could obtain an ROI and dramatically reduce the useless information for the following pathology segmentation.
Moreover, data normalization is also a vital data preprocessing technique to reduce data variations, due to the large domain shift among different modalities.
The majority of algorithms adopts the z-score normalization, while several teams simply scales the value to $[0, 1]$.

\textbf{Data augmentation.} 
Data augmentation schemes are widely used to facilitate model generalization ability.
The augmentation strategies utilized in the challenge methods could be roughly categorized into two parts, online transformation and offline data augmentation. 
Online transformation mainly contains several conventional augmentation schemes, such as randomly rotation, scaling, shifting, brightness, non-rigid transformations and contrast adjustment. 
For example,  USTB \cite{myops/USTB} transforms the original images non-rigidly with elastic-transform, grid-distortion and optical-distortion. 
%The performance of the non-rigid transformation is promising and has improved the Dice score of scar by almost 8\% on validation dataset.
Offline augmentation mainly refers to data generation. 
UBA \cite{myops/UBA} adopts spatially-adaptive normalization \cite{park2019semantic} for style transfer, pathology rotation and dilation/ erosion. 

\textbf{Model architectures.}
The design of model architecture is another determinant for the prediction results. 
The most prevailing backbone utilized by these methods is U-Net, which includes elegant symmetric encoder and decoder structure with multi-scale feature and skip-connection strategy. 
UESTC \cite{myops/UESTC} utilizes U-Net for both coarse and fine segmentation networks. 
Moreover, dense connection and attention are useful techniques. 
UBA \cite{myops/UBA} utilized an U-Net variant, termed as BCDU-Net \cite{tan2019efficientnet}, which reuses feature maps via dense convolutions.  
USTB \cite{myops/USTB} adopts channel and space attention modules at the basis of U-Net. 
%, and the champion structure of ISBI cell tracking challenge 2015, which extracts multi-scale features and leverages skip connection to maintain details. 
As for loss functions, most teams employ the Dice loss and the cross entropy loss. 
Also, FZU \cite{myops/FZU} utilizes boundary loss to enforce the model to focus on the boundary regions. 

\textbf{Post-processing.}
Among all teams, only four of them utilized post-processing to refine the predictions.
UBA \cite{myops/UBA} adopts the most complicated post-processing scheme. 
They first constrains the myocardium into a ring shape, and then calculates the distances of the pixels which are predicted as non background to the Myo. 
% Those pixels with distances less than the threshold were categorized into edema. Finally, they omitted the outliers, whose total volumes were less than 300 voxels. This refining strategy improved the Dice  of scar by almost 3\%. 
In addition, NPU \cite{conf/MyoPS/zhang2020} and USTB \cite{myops/USTB} simply removes the isolated regions. 
UOA \cite{conf/MyoPS/Kumaradevan2020} solely retains the largest connected component of predicted LV, and then fills the holes. 

\subsection{Data and results}
\subsubsection{Data}
The datasets of two challenges were collected from the same patient group, but with different annotations. 
The datasets includes multi-CMR sequences (bSSFP, T2 and LGE), taken from 45 subjects who underwent cardiomyopathy, and were annotated by at least three experts. 
They had been collected with institutional ethics approval and underwent anonymization. 
Moreover, multi-sequences were pre-aligned by MvMM algorithm \cite{journal/PAMI/zhuang2018} in the MyoPS challenge. 
The details of three sequences are shown in Table \ref{tb:table:dataset}.

\begin{table*}\center
\caption{Image acquisition parameters of the original multi-sequence data. ED: end-diastolic \cite{journal/MedIA/zhuang2020}.}
\label{tb:table:dataset}
{\scriptsize
% \begin{tabular}{ l|lllll *{7}{@{\ \,} l }} \hline
\begin{tabular}{  l| l l l l l*{7}{@{\ \,} l }}\hline
Sequence  & Imaging type  & No. slices & TR/ TE (ms) & Slice thickness & In-plane resolution\\
\hline
LGR CMR	  & T1-weighted              & 10-18	 & 3.6/1.8 & 5 mm    & 0.75 $\!\times\!$ 0.75 mm$^2$\\
T2 CMR    & T2-weighted & 3-7		 & 2000/90 & 12-20 mm &  1.35 $\!\times\!$ 1.35 mm$^2$ \\
bSSFP CMR & Cine sequence & 8-12		 & 2.7/1.4 & 8-13 mm  &  1.25 $\!\times\!$ 1.25 mm$^2$ \\
\hline
\end{tabular}}\\
\end{table*}

\subsubsection{Evaluation metrics} 
For evaluation, Dice score was applied in both MS-CMRSeg and MyoPS challenges,
\begin{align}
    Dice(V_{seg},V_{GD}) = \frac{2|V_{seg}\And V_{GD}|}{|V_{seg}|+|V_{GD}|},
\end{align}
where $V_{GD}$ and $V_{seg}$ denote the gold standard and automatic segmentation, respectively. 
Moreover, MS-CMRSeg used average surface distance (ASD) and Hausdorff distance (HD) as supplementary metrics, which can be defined as,
\begin{align}
    HD(X,Y) = max[\sup_{x\in X}\inf_{y\in Y}d(x,y),\sup_{y\in Y}\inf_{x\in X}d(x,y)],
\end{align}
\begin{equation}
    \text{ASD}(X, Y)=\frac{1}{2}\left(\frac{\sum_{x \in X} \min _{y \in Y}d(x, y)}{\sum_{x \in X} 1}+\frac{\sum_{y \in Y} \min _{x \in X}d(x, y)}{\sum_{y \in Y} 1}\right),
\end{equation}
where X and Y represent two sets of contour points, and $d(x, y)$ indicates the distance between the two points x and y.

\begin{figure*}[t]\center
    \includegraphics[width=0.98\textwidth]{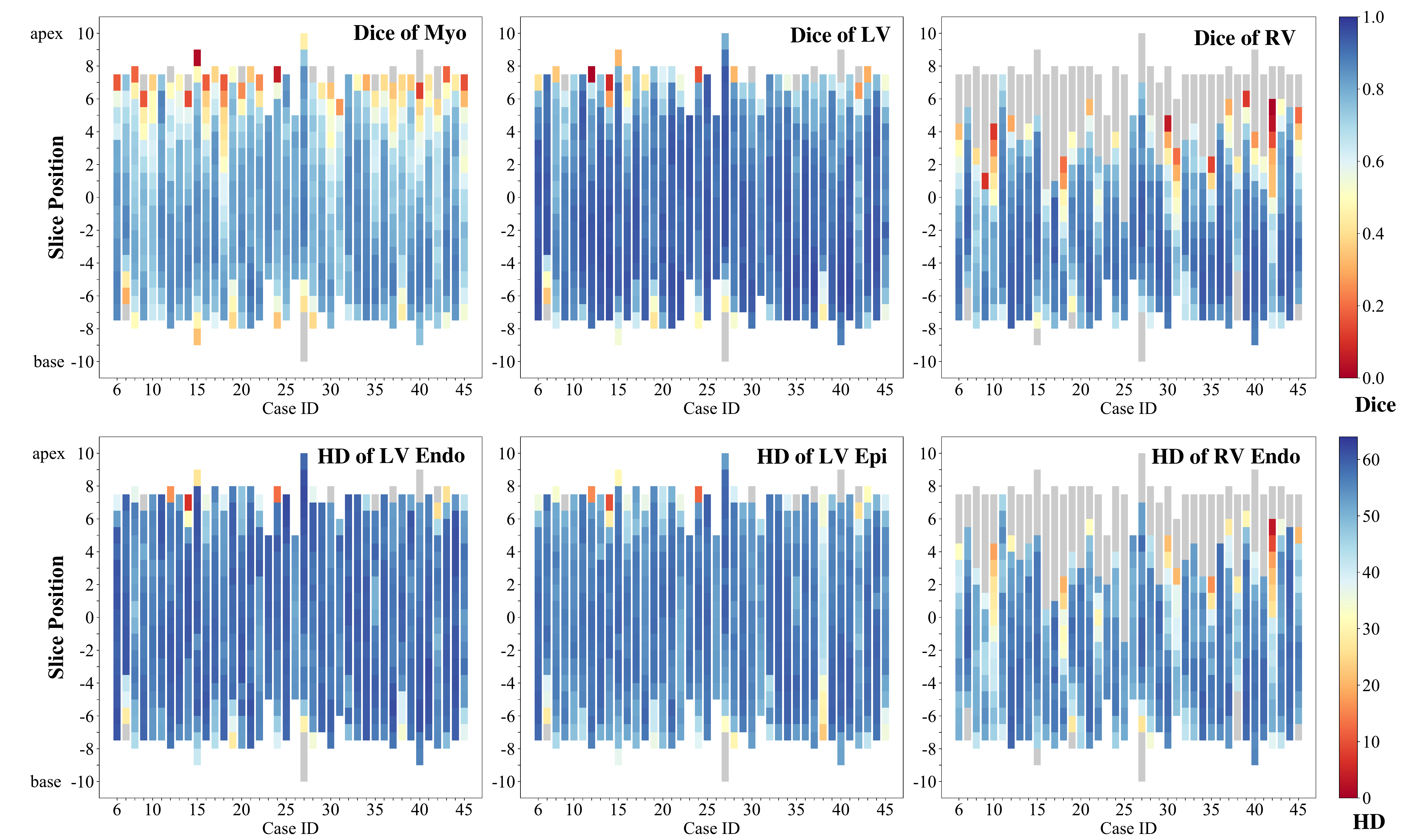}
  \caption{The average Dice score of the nine evaluated methods on each slice of the 40 test LGE CMR images. 
  Image adopted from Zhuang et al. \cite{journal/MedIA/zhuang2020}}.
\label{fig:MSCMR_spatial}
\end{figure*}

% Another thing worth mention is that the MvMM could be utilized as an unified registration workflow for MyoPS challenge data alignment. 
% As shown in Fig. \ref{fig:results_mvmm}, the per-alignment is a determinant for segmentation accuracy, that is the results of MvMM (a) outweighs those without registration (b). \Leicolor{}{(??, I don't get it)}

\subsubsection{Results from MS-CMRSeg challenge event}
Table \ref{tb:MSCMR result:Dice} and Table \ref{tb:MSCMR result:HD} present the quantitative results of the nine evaluated algorithms. 
The mean Dice scores of Myo, LV and RV segmentation were $0.766 \pm 0.104$, $0.891 \pm 0.056$ and $0.822 \pm 0.116$ respectively, and the mean volumetric HD values were $16.37 \pm 12.27$ mm (LV Endo), $18.06 \pm 12.18$ mm (LV Epi) and  $19.35 \pm 9.587$ mm respectively. 
Interestingly, the unsupervised domain adaptation models performed comparably to supervised ones, thanks to the style-translation data synthesis.  
As shown in Table \ref{tb:MSCMR result:Dice}, the Dice scores of LV were evidently better than that of RV and Myo.
Similarly, the HD values of myocardium including LV Endo and Epi, were generally better than that of RV Endo. 
One reasonable explanation is that the variance of the shape of RV is larger than that of LV, which commonly presents a ring-like structure. 
Moreover, the segmentation results were related to spatial position. 
Fig. \ref{fig:MSCMR_spatial} presents the accuracy of different slices.
One can see that the accuracy on apical slices were generally worse than that on midden and basal slices, which may be due to the shape variance.

\begin{table*} [t] \center
\caption{
Dice score of the evaluated algorithms on the LGE MRI segmentation \cite{journal/MedIA/zhuang2020}.
UDA: unsupervised domain adaptation;
Teams updating their results after the challenge deadline are indicated with an asterisk (*), and teams
using the unlabeled LGE images ($I^\mathrm{unl}$) for training are indicated with a dagger ($\dag$). %($I^\mathrm{unl}_{\mathrm{test}}$)
}
\label{tb:MSCMR result:Dice}
{\footnotesize
\begin{tabular}{l|lll|l}\hline
\multirow{2}*{Teams} & \multicolumn{3}{c|}{Volumetric Dice} & \multirow{2}*{Training}\\
\cline{2-4}
~ &  Myo & LV & RV & ~\\
\hline
% \begin{tabular}{l|lll|lll}\hline
% Teams &  Dice(Myo) & Dice(LV) & Dice(RV) &   HD(LV Endo)  &   HD(LV Epi) &  HD(RV Endo)\\
% \hline
ICL$^\dag$ \cite{conf/STACOM/chen2019}   &  $0.826 \pm 0.035$	& $0.919 \pm 0.026$ & $0.875 \pm 0.050$ & UDA\\
XMU$^\dag$ \cite{MSCMR/XIAN}  &  $0.796 \pm 0.059$	& $0.896 \pm 0.047$ & $0.846 \pm 0.086$ & UDA\\
INRIA$^*$\cite{MSCMR/INRIA}    &  $0.705 \pm 0.115$	& $0.870 \pm 0.051$ & $0.762 \pm 0.150$ & UDA\\
\hline
SCU$^*$\cite{conf/STACOM/wang2020}      &  $0.843 \pm 0.048$	& $0.926 \pm 0.028$ & $0.890 \pm 0.044$ &  Supervised\\
UB$^\dag$ \cite{MSCMR/UB}   &  $0.810 \pm 0.061$	& $0.898 \pm 0.045$ & $0.866 \pm 0.050$ &  Supervised\\
FAU	 \cite{mscmr/FAU}        &  $0.789 \pm 0.073$	& $0.912 \pm 0.034$ & $0.833 \pm 0.084$ &  Supervised\\
NVIDIA$^\dag$ \cite{mscmr/nvidia}&  $0.780 \pm 0.047$	& $0.890 \pm 0.043$ & $0.844 \pm 0.063$ &  Supervised\\
HIT$^\dag$ \cite{mscmr/HIT} &  $0.751 \pm 0.119$	& $0.884 \pm 0.070$ & $0.791 \pm 0.165$ &  Supervised\\
SUSTech	  \cite{MSCMR/SUST}  &  $0.610 \pm 0.102$	& $0.824 \pm 0.068$ & $0.710 \pm 0.135$ &  Supervised\\
\hline \hline
\multirow{3}*{Average} &  $0.775 \pm 0.093$ & $0.895 \pm 0.047$ & $0.828 \pm 0.114$ & UDA\\
~		 &  $0.764 \pm 0.109$ & $0.889 \pm 0.060$ & $0.822 \pm 0.117$ &  Supervised\\
~		 &  $0.766 \pm 0.104$ & $0.891 \pm 0.056$ & $0.822 \pm 0.116$ & All\\
\hline \hline
Inter-Ob   & $ 0.764 \pm 0.069 $ & $ 0.881 \pm 0.064 $&  $ 0.816 \pm 0.084$ &\\
\hline
\end{tabular} }\\
\end{table*}

% tb:result:DiceHD
\begin{table*} [t] \center
\caption{
HD of the evaluated algorithms on the LGE MRI segmentation \cite{journal/MedIA/zhuang2020}.
}
\label{tb:MSCMR result:HD}
{\footnotesize
\begin{tabular}{l|lll|l}\hline
\multirow{2}*{Teams} & \multicolumn{3}{c|}{Volumetric HD (mm)} & \multirow{2}*{Training}\\
\cline{2-4}
~ &  LV Endo &   LV Epi &  RV Endo & ~\\
\hline
ICL$^\dag$\cite{conf/STACOM/chen2019} & $10.28 \pm 3.376$	& $12.45 \pm 3.142$	& $15.38 \pm 6.942$ & UDA\\
XMU$^\dag$\cite{MSCMR/XIAN}  & $13.59 \pm 5.206$	& $15.70 \pm 5.814$	& $15.21 \pm 6.327$ & UDA\\
INRIA$^*$\cite{MSCMR/INRIA}   & $41.74 \pm 7.696$	& $42.79 \pm 13.26$	& $34.38 \pm 8.065$ & UDA\\
\hline
SCU$^*$\cite{conf/STACOM/wang2020}       & $9.748 \pm 3.280$	& $11.65 \pm 4.002$	& $13.34 \pm 4.615$ & Supervised \\
UB$^\dag$\cite{MSCMR/UB}     & $10.78 \pm 4.066$	& $11.96 \pm 3.620$	& $15.91 \pm 6.895$ &  Supervised\\
FAU	\cite{mscmr/FAU}          & $11.29 \pm 4.559$	& $12.54 \pm 3.379$	& $17.11 \pm 6.141$ &  Supervised\\
NVIDIA$^\dag$  \cite{mscmr/nvidia}& $11.58 \pm 7.524$	& $16.25 \pm 6.336$	& $18.12 \pm 9.262$ &  Supervised\\
HIT$^\dag$ \cite{mscmr/HIT}   & $14.30 \pm 8.170$	& $14.75 \pm 7.823$	& $17.87 \pm 9.322$ &  Supervised\\
SUSTech	 \cite{MSCMR/SUST}     & $23.69 \pm 14.66$	& $24.62 \pm 12.66$	& $23.46 \pm 7.596$ &  Supervised\\
\hline \hline
\multirow{3}*{Average}  & $21.87 \pm 15.23$ & $23.65 \pm 16.07$ & $21.66 \pm 11.49$ & UDA\\
~		  & $13.56 \pm 9.316$ & $15.30 \pm 8.389$ & $17.64 \pm 8.092$ &  Supervised\\
~		  & $16.37 \pm 12.27$ & $18.06 \pm 12.18$ & $19.35 \pm 9.587$ & All\\
\hline \hline
Inter-Ob  & $ 12.03 \pm 4.443 $ & $ 14.32 \pm 5.164 $&  $ 21.53 \pm 9.460 $ & \\
\hline
\end{tabular} }\\
\end{table*}

% \begin{figure*}[t]\center
%     \includegraphics[width=\textwidth]{img/MyoPSimgs/results for MyoPS.png}
%   \caption{a.The correlation between the Dice and the per- centage of scar/edema.b.Bulls-eye plots of the regional pathology analysis (AHA 18-segment model) obtained by averaging results of all test subjects. The regional performance of scar and edema segmentation in terms of $Spe^*$.c.The average $Spe^*$ of both scar and edema segmentation by the fifteen evaluated methods on each slice of the 20 test images. Here, average Dice and HD values of the nine evaluated methods on each slice of the 40 test LGE CMR images. Image adopted from Zhuang et al. \cite{}. \Leicolor{}{Hi Sihan, note that these images are still not published. Once we submit our MyoPS benchmark paper, we can cite it later.}}
% \label{fig:MSCMR}
% \end{figure*}

\subsubsection{Results from MyoPS challenge event}
% Table \ref{tb:result:scar and edema} presents the MyoPS results of all the evaluated algorithms.
The mean Dice scores of the scar and edema were $0.634 \pm 0.225$ and $0.665 \pm 0.146$ respectively, and the best Dice scores for both scar and edema were achieved by ESTC.  %上面MSCMR用一般现在时，这里又用过去式？一般描述实验结果都是用过去式，方法都用一般现在时。所以其实上面描述challenge方法也可以改成一般现在时
The results indicated that the Dice scores of edema are higher than that of scars, maybe due to the larger extent percentage of edema. 
Moreover, the segmentation precise is also related with slice position, as we found that the accuracy of midden and basal slices are relatively higher than that of apical slices.
One possible reason is that the large variation of both anatomical and pathological shapes on apical slices. 
To figure out the correlation between the accuracy and pathology regional position, we also generated the bulls-eye maps to visualize the segmentation accuracy per region. 
The maps show that the inaccurate segmentation of scars and edema mainly occurs at the basal and inferior regions.

\subsection{Discussion and conclusion}
Intrinsically, the aforementioned works are trying to explore and analyze the same dataset progressively. 
Firstly, the inherent challenge for the multi-sequence task is that the images obtained from different sequences or even from the same center are generally not well-aligned. 
%In this case, MvMM was firstly used for automatic intra- and inter-image registration, which will be beneficial for following segmentation. 
MS-CMRSeg challenge focused on the anatomical structure segmentation, while the MyoPS challenge was an extent of the MS-CMRSeg for the pathology segmentation.
Observed from the evaluated methods, scar and edema segmentation will be benefited from cardiac structure information, hence the majority of methods leveraged ``coarse to fine" strategy. 
Note that the MS-CMRSeg and MyoPS dataset and corresponding evaluation tool are still available for researchers upon request and registration via the challenge homepage \textit{https://zmiclab.github.io/index.html}{https://zmiclab.github.io/index.html}.

\section{LGE MRI based left atrial scar segmentation and quantification}\label{LA}
\subsection{Introduction}
\begin{figure*}[t]\center
    \includegraphics[width=1\textwidth]{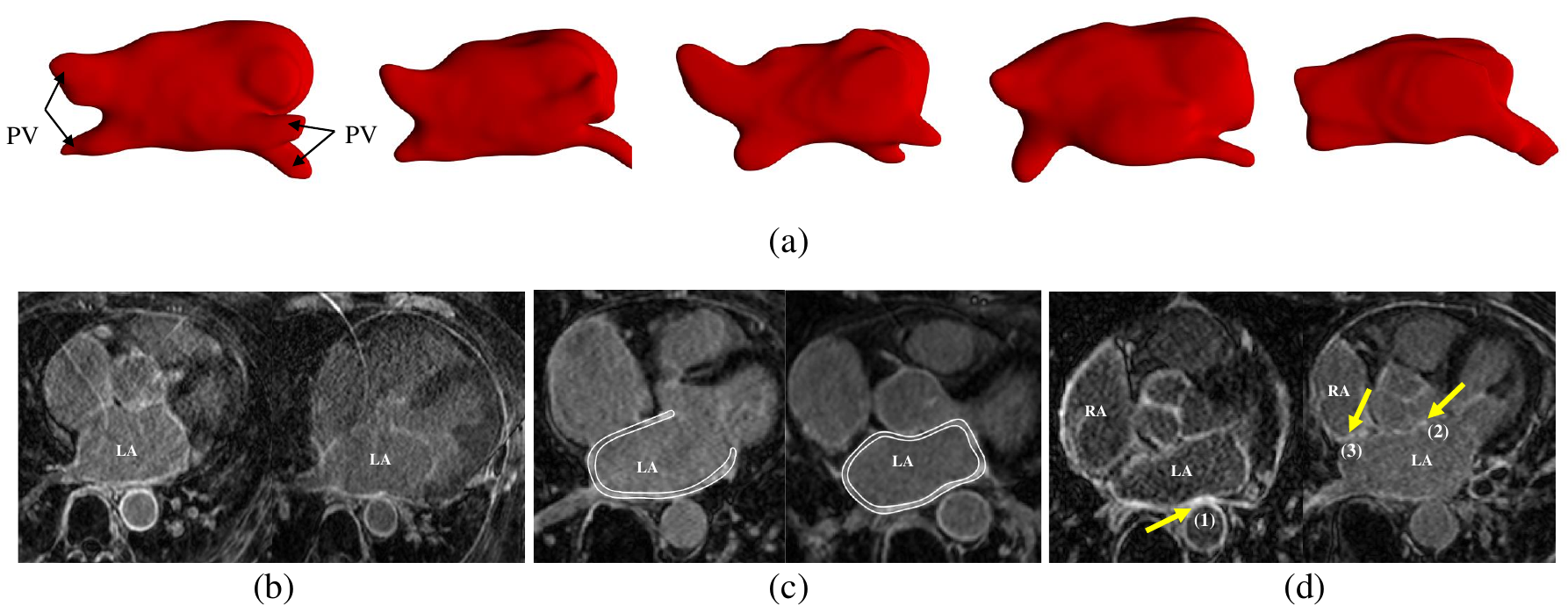}
  \caption{The challenges of automatic scar quantification from LGE MRI:
     (a) various LA and pulmonary vein (PV) shapes;
     (b) two typical LGE MRIs with poor quality;
     (c) thin atrial walls highlighted using bright white color;
     (d) surrounding enhanced regions pointed out by the arrows, where (1) and (2) indicate the enhanced walls of descending and ascending aorta, respectively; and (3) denotes the enhanced walls of right atrium.
     Images adopted from Li et al. \cite{journal/MedIA/li2022}.}
\label{fig:LA intro:challenges}
\end{figure*}

Atrial fibrillation (AF) is the most common type of arrhythmia and a severe public health concern.
The identification of fibrosis and scarring region in left atrium (LA) is important for AF diagnosis, treatment and prognosis.
Late gadolinium enhancement magnetic resonance imaging (LGE MRI) has been shown to be an non-invasive tool for left atrial (LA) fibrosis and scar assessment and quantification.
However, LGE MRI usually has poor image quality, mainly due to its residual respiratory motion, variability in the heart rate and gadolinium wash-out during the long acquisition time.
Moreover, LA wall is thin (mean thickness: $1.89\pm 0.48~mm$ \cite{journal/JCE/beinart2011}) with regional wall thickness variations.
Various shapes of LA and pulmonary veins (PV) introduce additional challenges for scar segmentation of LGE MRI.
Figure \ref{fig:LA intro:challenges} visualizes the major challenges.

Most methods for scar segmentation and quantification mainly based on thresholding \cite{journal/EP/pontecorboli2017}. 
Recently, with the development of deep learning (DL) in medical image computing, several DL-based algorithms have been proposed for automatic scar segmentation and quantification \cite{journal/MedIA/li2020,journal/MedIA/li2021}. 
One could refer to the review papers \cite{journal/EP/pontecorboli2017,journal/MedIA/li2022} for the literature of LA and scar segmentation from LGE MRI.
Next, we will introduce our two state-of-the-art framework for LA LGE MRI computing.
Our first work mainly aim to solve the challenging of thin thickness in scar quantification, and we proposed a surface projection scheme (see Section \ref{LearnGC}).
The second work focuses on the joint optimization of LA segmentation and scar quantification, and we designed a multi-task learning based network (see Section \ref{AtrialJSQnet}).

% \begin{figure*}[t]\center
%     \subfigure[] {\includegraphics[width=1\textwidth]{fig_intro_LAshape}}
%     \subfigure[] {\includegraphics[width=0.32\textwidth]{fig_intro_LGE-MRI}}
%     \subfigure[] {\includegraphics[width=0.32\textwidth]{fig_intro_LAwall}}
%     \subfigure[] {\includegraphics[width=0.32\textwidth]{fig_intro_noise}}
%   \caption{The challenges of automatic segmentation and quantification of LGE MRI for AF:
%      (a) various LA and pulmonary vein (PV) shapes;
%      (b) two typical LGE MRIs with poor quality;
%      (c) thin atrial walls highlighted using bright white color in the figure;
%      (d) surrounding enhanced regions pointed out by the arrows, where (1) and (2) indicate the enhanced walls of descending and ascending aorta, respectively; and (3) denotes the enhanced walls of right atrium (RA).
%      Images (b)-(d) adopted from \citet{journal/MedIA/li2020}.}
% \label{fig:LA intro:challenges}
% \end{figure*}

\subsection{Method}
\subsubsection{LearnGC: Atrial Scar Segmentation via Potential Learning in the Graph-Cut Framework} \label{LearnGC}
\begin{figure*}[t]\center
    \includegraphics[width=0.98\textwidth]{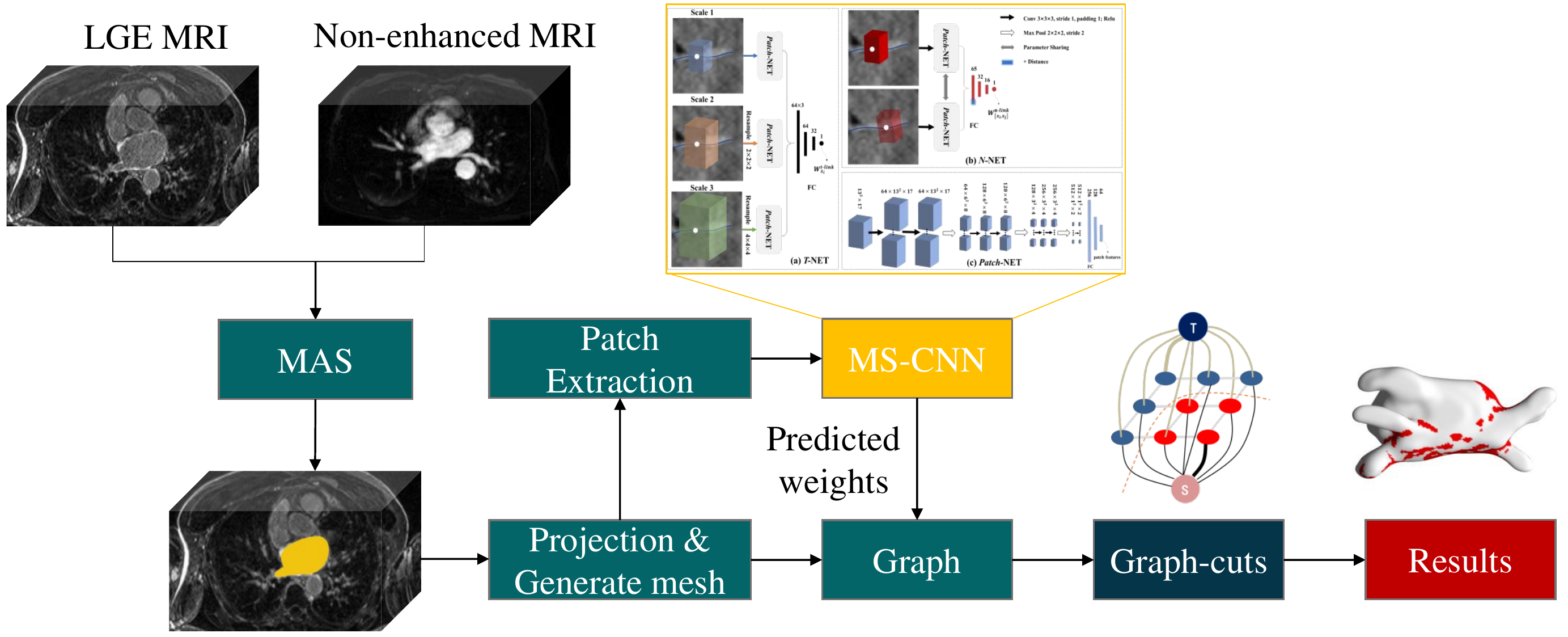}
  \caption{The proposed LearnGC framework for the scar quantification. The images modified from Li et al. \cite{journal/MedIA/li2020,journal/MedIA/li2021}.}
\label{fig:method:LearnGC}
\end{figure*}

Figure \ref{fig:method:LearnGC} presents the proposed LearnGC framework.
First, we employ the multi-atlas segmentation (MAS) algorithm for whole heart segmentation, and then extract LA segmentation as an initialization.
Then, we generate a surface mesh based on the LA endocardium to perform the scar quantification on it via graph-cuts algorithm.
The edge weights of graph-cuts are predicted by the proposed multi-scale convolution network (MS-CNN).
Note that the scar quantification is performed on the generated surface mesh, to avoid the challenging segmentation task, i.e., thin LA wall segmentation.
Moreover, it could be effective to reduce the computational cost.

\textbf{LA segmentation via MAS.}
As LGE MRI covers the whole heart, we propose to employ multi-atlas whole heart segmentation (MA-WHS) to obtain the geometrical information of the LA.
MAS algorithm can be separated into two steps: registration between atlases and the target image; label fusion of transformed atlases.
Considering the poor image quality of LGE MRI, we employ an non-enhanced MRI from the same patient to assist its segmentation.
The non-enhanced MRI normally has higher quality, and shares the anatomical structures with LGE MRI.
With the WHS results, we employ marching cube algorithm to obtain a surface mesh of the LA endocardium.

\textbf{Projection and graph formulation.}
In clinic, the location and extent of scars are the main concern.
Inspired by this, we project the LA endocardium onto a surface mesh, and then performed the scar quantification on the surface via graph-cuts.
Moreover, the projection mitigates the effect of LA wall thickness and inaccurate LA segmentation, and also reduces dramatically the computational complexity.
The projection is equidistant to preserve the geodesic distances between two nodes in the graph-cuts framework.
Instead of using single pixel on the surface, we extract a profile via multi-scale patch to incorporate both global and local texture information of scars.

The scar quantification is formulated as an energy minimization problem in the graph-cuts framework.
The edge weights of graph include the regional term $E_{R}$ and the boundary term $E_{B}$, which encodes the intensity distributions of two classes and ensures the continuity between neighbors.
One can denote the graph as $G=\{\mathcal{X},\mathcal{N}\}$, where $\mathcal{X}=\{x_i\}$ is the set of graph nodes and $\mathcal{N}=\{<x_i,x_j>\}$ indicates the set of edges connecting graph nodes. 
Each graph has two terminals, which denote the scars (foreground) and normal myocardium (background), respectively.
There are two kinds of edges, i.e., t-link that connects graph nodes to the terminals and n-link connecting neighboring nodes
\cite{conf/iccv/Boykov2001}.
Therefore, the segmentation energy can be defined as follows,
\begin{equation}
\begin{array}{l@{\ }l}
 E(l) &= E_R(l) + \lambda E_B(l) \\
 &=\displaystyle\sum_{x_i\in \mathcal{X}}  W_{x_i}^{\textit{t-link}}(l_{x_i})
 +\lambda \sum_{(x_i, x_j)\in \mathcal{N}}  W_{\{x_i, x_j\}}^{\textit{n-link}}(l_{x_i}, l_{x_j}),
\end{array}
\label{method:LearnGC:energy func}
\end{equation}
where $W_{x_i}^{\textit{t-link}}$ and $W_{\{x_i, x_j\}}^{\textit{n-link}}$ are the t-link and n-link weight, respectively;
$l_{x_i}\in\{0,1\}$ is the label value of $x_i$;
and $\lambda$ is a balancing parameter.
Different from conventional graph-based segmentation, we directly predict the t/n-link weights for the regional and boundary terms.
In this way, we can represent each graph node using a multi-scale patch (MSP), and then learn the weighs using the proposed MS-CNN.
These patches are defined along the normal direction of the LA endocardial surface with an elongate shape.
We further adjust the sample spacing to generate MSPs on the LGE MRI with different resolutions.

\textbf{Edge weight prediction via MS-CNN.}
To predict the edge weights in Equation \ref{method:LearnGC:energy func}, we design two networks, i.e., $T$-NET and $N$-NET.
$T$-NET is designed to predict the t-link weights, i.e., the probabilities of a node belonging to scarring and normal regions respectively;
$N$-NET aims to calculate the n-link weights, defined based on the similarity of two neighbour nodes and their distance.
To embed the MSP into the networks, we adopt parallel convolutional pathways for multi-scale training, namely MS-CNN.
In the training phase of MS-CNN, we adopt a weight sampling strategy to mitigate the class imbalance problem.
Specifically, instead of extracting the patches of all nodes for training, we randomly select the similar number of nodes from the normal wall and scars. 
Besides, we use a random shift strategy when extracting the MSPs to mitigate the effect of inaccurate LA segmentation.
Note that the weight sampling and random shift strategy are not required in the testing phase.
After obtain the weights, one can obtain the scar quantification result by optimizing the graph-cuts framework.

\subsubsection{AtrialJSQnet: A New Framework for Joint Segmentation and Quantification of Left Atrium and Scars Incorporating Spatial and Shape Information} \label{AtrialJSQnet}
\begin{figure*}[t]\center
    \includegraphics[width=0.98\textwidth]{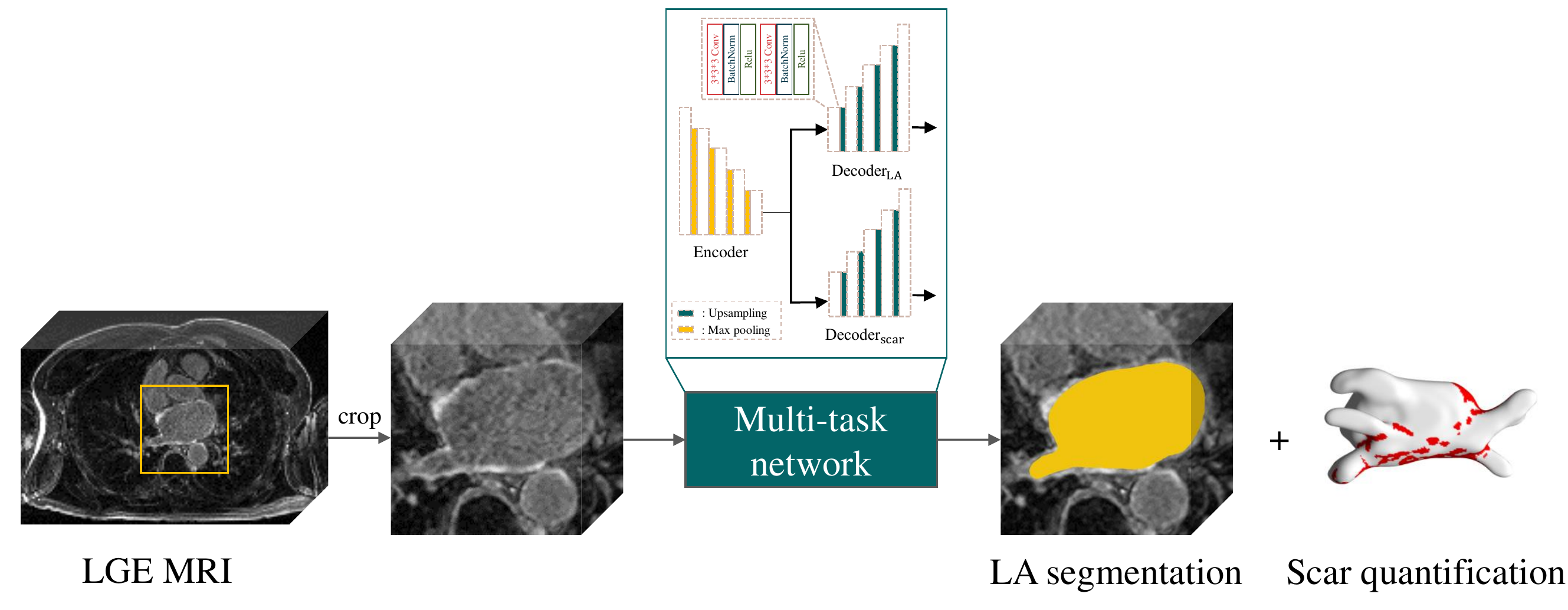}
  \caption{The proposed AtrialJSQnet framework for the simultaneous LA segmentation, scar projection onto LA surface, and scar quantification. The images modified from Li et al. \cite{journal/MedIA/li2021}.}
\label{fig:method:AtrialJSQnet}
\end{figure*}

Figure \ref{fig:method:AtrialJSQnet} presents the overview of the proposed joint segmentation and quantification framework, i.e., AtrialJSQnet.
One can see that AtrialJSQnet is a two-task network consisting of two decoders for LA segmentation and scar quantification, respectively.
We proposed a spatial encoding (SE) loss as to learn the spatial information of the LA cavity and scars.
Moreover, we employ a shape attention (SA) scheme to utilize the spatial relationship between LA and scar.
The SA scheme is also helpful to achieve an end-to-end scar projection.

\textbf{Spatially encoded constraint in the AtrialJSQnet framework.}
We introduce a novel SE loss, which incorporates spatial information in the pipeline without any modifications of networks. The SE loss is designed based on the distance transform maps (DTM), a continuous representation of the target label.
The signed DTM can be defined as,
\begin{equation}
\phi(x_i)=\begin{cases} -d^\beta & x_i \in \Omega_{in} \\0 & x_i \in S\\ d^\beta & x_i \in \Omega_{out} \end{cases}
\label{eq:SDM}
\end{equation}
where $\Omega_{in}$ and $\Omega_{out}$ indicate the region inside and outside the target label, respectively;
$S$ denotes the surface boundary, $d$ represents the distance from pixel $x_i$ to the nearest point on $S$, and $\beta$ is a hyperparameter.

For LA segmentation, the SE loss is defined as,
\begin{equation}
  \mathcal L_{LA}^{SE} = \sum_{i=1}^N (\hat{y}(x_i; \theta)-T_{LA}) \cdot \phi(x_i),
\end{equation}
where $\hat{y}$ and $y$ ($y\in\{0,1\}$) are the prediction of LA and its ground truth, respectively;
$N$ is the number of pixels, $T_{LA}$ is the threshold for LA segmentation, and $\cdot$ denotes the dot product.
One can see the main idea of $\mathcal L_{LA}^{SE}$ is to assign different penalties to false classification of each pixel based on its distance to the target boundary, i.e., the DTM value.
The final loss for LA segmentation can be defined as,
\begin{equation}
  \mathcal L_{LA} = \mathcal L_{LA}^{BCE} + \lambda_{LA}\mathcal L_{LA}^{SE},
\end{equation}
where $\mathcal L_{LA}^{BCE}$ is the conventional binary cross entropy (BCE) loss, and $\lambda_{LA}$ is a balancing parameter.

For scar quantification, we first obtain the DTM of scars and normal wall, and then calculate the corresponding probability maps based on DTMs, i.e., $p=exp^{-|\phi(x)|}$ and $p= [p_{normal}, p_{scar}]$.
Therefore, the SE loss for scar quantification is defined as,
\begin{equation}
  \mathcal L_{scar}^{SE} = \sum_{i=1}^N \|\hat{p}(x_i; \theta) - p(x_i)\|^2_2,
\label{eq:SE_scar}
\end{equation}
where $\hat{p}$ ($\hat{p} = [\hat{p}_{normal}, \hat{p}_{scar}]$) is the predicted probability maps of normal wall and scars.
Note that here we did not consider the probability map of background (pixels not belong to scars neither normal wall), as we quantify the scars on the LA surface.
However, the predicted LA surface can be inaccurate, so we did not employ a fixed threshold on $\hat{p}_{normal}$ or $\hat{p}_{scar}$.
Instead, we propose to compare the probabilities of each pixel belonging to scars and normal wall for final scar quantification results.

\textbf{End-to-end trainable shape attention via multi-task learning.}
For joint optimization, we develop a multi-task network where segmentation and quantification of LGE MRI are simultaneously achieved.
To further learn the spatial relationship between LA and scars, we introduce an attention mask, which is represented by the LA boundary from the predicted LA segmentation.
The attention mask not only can alleviate the class imbalance problem, but also contributes to the end-to-end surface projection.
To achieve this, we design a SA loss which is defined as follows,
\begin{equation}
  \mathcal L^{SA}_{scar} = \sum_{i=1}^N (M \cdot (\nabla \hat{p}(x_i; \theta) - \nabla p(x_i)))^2,
\end{equation}
where $\nabla \hat{p} = \hat{p}_{normal} - \hat{p}_{scar}$, 
$\nabla p = p_{normal}-p_{scar}$, and $M=\left\{M_{1}, M_{2}\right\}$ is the attention mask.
Here, $M_{1}$ and $M_{2}$ refer to the gold standard LA wall segmentation and the predicted LA boundary, respectively.

Therefore, we can combine these loss functions for the final optimization of AtrialJSQnet as follows,
\begin{equation}
   \mathcal L = \mathcal L_{LA} + \lambda_{scar}\mathcal L_{scar}^{SE}
        + \lambda_{M_{1}}\mathcal L^{SA}_{scarM_1} + \lambda_{M_{2}}\mathcal L^{SA}_{scarM_2},
\end{equation}
where $\lambda_{scar}$, $\lambda_{M_{1}}$ and $\lambda_{M_{2}}$ are balancing parameters.

\subsection{Data and results}
\subsubsection{Data acquisition}
For the proposed methods, we evaluated them via two different dataset, separately.
For LearnGC, we employed 58 post-ablation LGE MRIs from longstanding persistent AF patients.
All images were acquired from 1.5T Siemens Magnetom Avanto scanner (Siemens Medical Systems, Erlangen, Germany), using an inversion prepared segmented gradient echo sequence (TE/TR 2.2/5.2 ms), and were acquired 15 min after gadolinium administration.
The acquisition resolution is (1.4-1.5) $\!\times\!$ (1.4-1.5) $\!\times\!$ 4 mm, and is reconstructed to (0.7-0.75) $\!\times\!$ (0.7-0.75) $\!\times\!$ 2 mm.
For AtrialJSQnet, we adopted a public data from the \textit{MICCAI2018 Atrial Segmentation Challenge} \cite{link/LAseg2018}, which released 60 post-ablation LGE MRIs with manual LA segmentation results.
The public LGE MRIs were acquired from 1.5T Siemens Avanto or 3T Siemens Vario (Siemens Medical Solutions, Erlangen, Germany), with a resolution of 1.25 $\!\times\!$ 1.25 $\!\times\!$ 2.5 mm.
All images were acquired approximately 20–25 min after gadolinium administration, using a 3D respiratory navigated, inversion recovery prepared gradient echo pulse sequence (TE/TR 2.3/5.4 ms). 

\subsubsection{Gold standard and evaluation}
All the images were manually segmented by a well-trained expert, which were regarded as the gold standard for evaluation.
For LA segmentation, we employed Dice score, ASD and HD for evaluation.
% \begin{equation}
%   \text{Dice}(V_{\text{auto}}, V_{\text{manual}}) = \frac{2\left|V_{\text{auto}} \cap V_{\text{manual}}\right|}{\left|V_{\text{auto}}\right|+\left|V_{\text{manual}}\right|},
% \end{equation}
% \begin{equation}
%     \text{ASD}(X, Y)=\frac{1}{2}\left(\frac{\sum_{x \in X} \min _{y \in Y}d(x, y)}{\sum_{x \in X} 1}+\frac{\sum_{y \in Y} \min _{x \in X}d(x, y)}{\sum_{y \in Y} 1}\right),
% \end{equation}
% and
% \begin{equation}
%   \text{HD}(X, Y)=\max \Big[\sup _{x \in X} \inf _{y \in Y} d(x, y), \sup _{y \in Y} \inf _{x \in X} d(x, y)\Big],
% \end{equation}
% where $V_{\text{manual}}$ and $V_{\text{auto}}$ denote the manual and automatic segmentation, respectively;
% $X$ and $Y$ represent two sets of contour points;
% $d(x, y)$ indicates the Euclidean distance between the two points $x$ and $y$;
% and $|\cdot|$ means the number of pixels.
For scar quantification evaluation, we firstly projected manual and (semi-) automatic scar segmentation results onto the surface of manual LA segmentation.
Then, we used the \textit{Accuracy}, Dice score and generalized Dice (GDice) score for the evaluation of scar quantification.
\begin{equation}
    \textit{Accuracy}=\frac{TP+TN}{TP+FP+FN+TN}, 
\end{equation}
where $TP$, $TN$, $FN$ and $FP$ stand for the number of true positives, true negatives, false negatives and false positives, respectively.
\begin{equation}
  \text{GDice} = \frac {2\sum_{k=0}^{N_{k}-1}\left| {S}_{k}^{\textit{auto}} \cap {S}_{k}^{\textit{manual}}\right|} {\sum_{k=0}^{N_{k}-1}(\left| S_{k}^{\textit{auto}}\right| + \left|S_{k}^{\textit{manual}}\right|)},
\end{equation}
where $S_{k}^{\textit{auto}}$ and $S_{k}^{\textit{manual}}$ indicate the segmentation results of label $k$ from the automatic method and manual delineation on the LA surface, respectively, $N_{k}$ is the number of labels, and $N_{k}=2$ here to represent scarring ($k=1$) and normal wall ($k=0$) regions.

\subsubsection{Performance of the proposed method}

Here, we will present the results for testing LearnGC and AtrialJSQnet algorithms, respectively.

\textbf{LearnGC.}
Table \ref{tb:LearnGC result:scar} presents the LearnGC quantification results, which includes the results of both comparison and ablation study.
The proposed method is LA$_\text{auto}$ + LearnGC, where the LA segmentation was performed via MA-WHS, the weights of the graph were predicted using MS-CNN, and the balancing parameter $\lambda$ was set to 0.4.
One can see that proposed LearnGC method obtained the best performance compared to other comparison methods.
It indicated that the proposed method can obtain the state-of-the-art performance, and also proved the effect of the proposed schemes, such as MS-CNN, random shift, and graph-cuts.

\begin{figure}[tb]\center
 \includegraphics[width=0.65\textwidth]{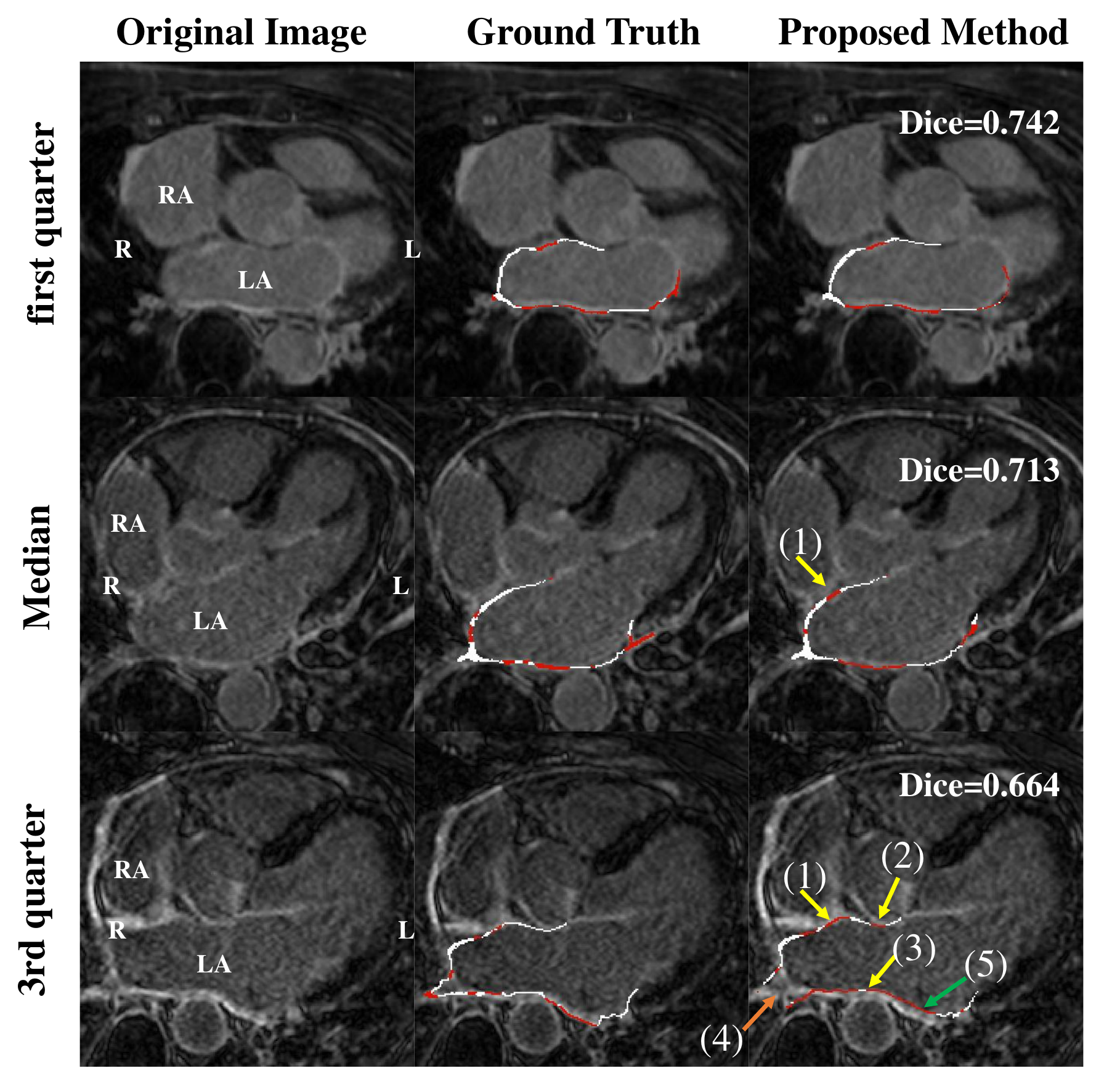} 
  \caption{Axial view of the images, the ground truth scar segmentation and the results by the proposed method. The red and white color labels represent the scar and normal wall, respectively. 
  Arrow (1), (2) and (3) indicate the major classification errors of the proposed method caused by the surrounding enhanced regions, respectively from the right atrium wall, ascending aorta wall and descending aorta wall; 
  arrow (4) shows an error from the misalignment between the automatic LA segmentation and the ground truth; 
  arrow (5) illustrates that the proposed method can still perform well, even though the automatic LA segmentation contains obvious errors.}
\label{fig:LearnGC result:2dvisual}
\end{figure}

Figure \ref{fig:LearnGC result:2dvisual} presents the 3D visualization of scar quantification results by the nine methods.
One can see that the 3D visualization results were consistent with the above quantitative analysis.
The predicted scars based on LA$_\text{M}$ and LA$_\text{auto}$ were projected onto two surfaces, i.e., GT$_\text{M}$ and GT$_\text{auto}$.
However, one can see that GT$_\text{M}$ and GT$_\text{auto}$ have similar scar distribution.
Compared to other methods, the proposed LearnGC obtained the most accurate and smooth scar quantification results.
Figure \ref{fig:LearnGC result:2dvisual} visualizes the axial view of three representative cases.
The illustration further proved that the proposed method could obtain promising scar quantification results, though with some minor errors.
The mis-classification indicates the major challenges of automatic scar quantification, contributed to the major errors of scar quantification.

\begin{table*} [t] \center
    \caption{
    Summary of the quantitative evaluation results of scar quantification. 
    Here, LA$_\text{M}$ denotes the manual LA segmentation, while LA$_\text{auto}$ refers to the automatically segmented LA using MA-WHS.
    $^0$ means that the methods did not employ random shift scheme ($\gamma$=0).
    MS-CNN refers to the learning based method only using the two t-link weights estimated from $T$-NET to classify scars.
    The asterisk ($^*$) in column Dice$_\text{scar}$ indicates the methods performed statistically poorer ($p<0.01$)
	compared to the proposed LA$_\text{auto}$ + LearnGC. 
     }
\label{tb:LearnGC result:scar}
{\small
\begin{tabular}{ l| l *{3}{@{\ \,} l }}\hline
Method       & \quad Accuracy & \qquad Dice$_\text{scar}$  & \qquad $G$Dice \\
\hline
LA$_\text{M}$ + 2SD \cite{journal/JCMR/Karim2013}  &$ 0.809 \pm 0.074 $&  \quad $ 0.275 \pm 0.091^* $&  \quad $ 0.758 \pm 0.098 $  \\
LA$_\text{M}$ + Otsu \cite{journal/TSMC/Otsu1979}           &$ 0.763 \pm 0.188 $&  \quad $ 0.396 \pm 0.090^* $&  \quad $ 0.726 \pm 0.207 $  \\
LA$_\text{M}$ + MGMM \cite{journal/TBME/liu2017}           &$ 0.708 \pm 0.160 $&  \quad $ 0.545 \pm 0.101^* $&  \quad $ 0.716 \pm 0.190 $  \\
LA$_\text{M}$ + MGMM + GC       &$ 0.716 \pm 0.162 $&  \quad $ 0.562 \pm 0.102^* $&  \quad $ 0.721 \pm 0.192 $  \\
\hline
LA$_\text{M}$ + U-Net \cite{conf/MICCAI/ronneberger2015}          &$ 0.832 \pm 0.046 $&  \quad $ 0.568 \pm 0.083^* $&  \quad $ 0.826 \pm 0.052 $  \\
LA$_\text{M}$ + MS-CNN$^0$     &$ 0.798 \pm 0.051 $&  \quad $ 0.615 \pm 0.083^* $&  \quad $ 0.811 \pm 0.047 $  \\
\hline\hline
LA$_\text{auto}$ + MS-CNN$^0$  &$ 0.806 \pm 0.052 $&  \quad $ 0.631 \pm 0.080^* $&  \quad $ 0.814 \pm 0.047 $  \\
LA$_\text{auto}$ + MS-CNN      &$ 0.846 \pm 0.032 $&  \quad $ 0.692 \pm 0.069^* $&  \quad $ 0.851 \pm 0.030 $  \\
LA$_\text{auto}$ + LearnGC     &$ 0.856 \pm 0.033 $&  \quad $ 0.702 \pm 0.071   $&  \quad $ 0.859 \pm 0.031 $  \\
\hline
\end{tabular} }\\
\end{table*}

\begin{figure}[t]\center
  \includegraphics[width=0.75\textwidth]{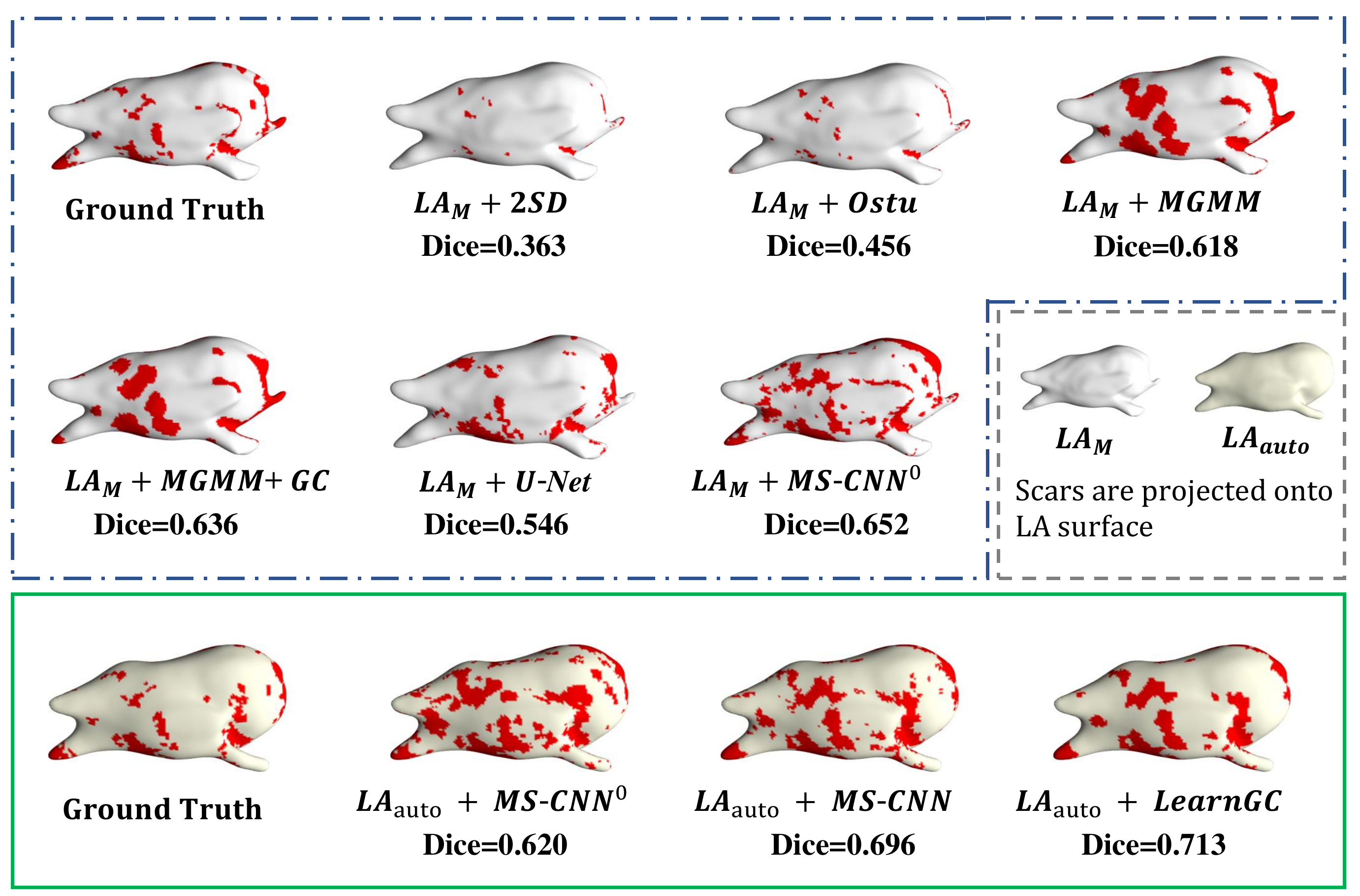}
  \caption{3D visualization of the LA scar classification results using the nine methods.
  This is the median case selected from the test set in terms of Dice score of scars by the proposed method.
  The scarring areas are red-colored on the LA surface mesh, which can be constructed either from $LA_M$ (LA surface in white) or from $LA_{auto}$ (LA surface in light yellow). Image adopted from Li et al. \cite{journal/MedIA/li2020}}
\label{fig:LearnGC result:3dvisual}
\end{figure}

\textbf{AtrialJSQnet.}
Table \ref{tb:AJSQnet result:LA} and Table \ref{tb:AJSQnet result:scar} present the LA segmentation and scar quantification results of different methods, respectively.
For the LA segmentation, the proposed SE loss performed better than the conventional losses, such as BCE and Dice loss.
For the scar quantification, the SE loss also obtained promising performance compared to the conventional losses in terms of Dice$_\mathrm{scar}$.
The three (semi-) automatic scar quantification methods generally obtained acceptable results, but relied on an accurate initialization of LA.
LearnGC had a similar result compared to MGMM in Dice$_\mathrm{scar}$, but its \emph{Accuracy} and $G$Dice were higher.
The proposed method performed statistically significant better than all the automatic methods in terms of Dice$_\mathrm{scar}$ ($p\leq0.001$).
Both the LA segmentation and scar quantification benefited from the proposed joint optimization scheme comparing to separately optimize the two tasks.
After introducing the new SA loss, the results were further improved in terms of Dice$_\mathrm{scar}$ ($p\leq0.001$), but with a slightly worse \emph{Accuracy} ($p\leq0.001$) and $G$Dice ($p>0.1$).
Moreover, with the SA loss some small and discrete scars could be detected, and an end-to-end scar quantification and projection were achieved.

\begin{table*} [t] \center
    \caption{
    Summary of the quantitative evaluation results of LA segmentation. Here, U-Net$_\text{LA}$ uses the original U-Net architecture for LA segmentation; 
    BCE, SE, SA and SESA refer to the different loss functions.
    The proposed method is denoted as AJSQnet-SESA.}
\label{tb:AJSQnet result:LA}
{\small
\begin{tabular}{ l| l *{5}{@{\ \,} l }}\hline
Method    & \quad Dice$_\text{LA}$  & \quad ASD (mm) & \quad HD (mm) \\
\hline
U-Net$_\text{LA}$-BCE    &$ 0.889 \pm 0.035 $&  \quad $ 2.12 \pm 0.797 $& \quad $ 36.4 \pm 23.6 $\\
U-Net$_\text{LA}$-Dice   &$ 0.891 \pm 0.049 $&  \quad $ 2.14 \pm 0.888 $& \quad $ 35.0 \pm 17.7 $\\
U-Net$_\text{LA}$-SE     &$ 0.880 \pm 0.058 $&  \quad $ 2.36 \pm 1.49  $& \quad $ 25.1 \pm 11.9 $\\
\hline\hline
AJSQnet-BCE                &$ 0.890 \pm 0.042 $&  \quad $ 2.11 \pm 1.01 $&  \quad $ 28.5 \pm 14.0 $\\
AJSQnet-SE                 &$ 0.909 \pm 0.033 $&  \quad $ 1.69 \pm 0.688 $&  \quad $ 22.4 \pm 9.80 $\\
AJSQnet-SESA               &$ 0.913 \pm 0.032 $&  \quad $ 1.60 \pm 0.717 $&  \quad $ 20.0 \pm 9.59 $\\
\hline \hline
Inter-ob                   &$ 0.894 \pm 0.011 $&  \quad $ 1.81 \pm 0.272$&  \quad $ 17.0 \pm 5.50 $\\
\hline
\end{tabular} }\\
\end{table*}

\begin{table*} [t] \center
    \caption{                                                                                                         
    Summary of the quantitative evaluation results of scar quantification. 
    Here, LA$_\text{M}$ denotes that scar quantification is based on the manually segmented LA, while LA$_{\text{U\mbox{-}Net}}$ indicates that it is based on the U-Net$_\text{LA}$-BCE segmentation;
    U-Net$_\text{scar}$ is the scar segmentation directly based on the U-Net architecture with different loss functions; 
    The inter-observer variation (Inter-Ob) is calculated from randomly selected twelve subjects.
     }
\label{tb:AJSQnet result:scar}
{\small
\begin{tabular}{ l| l *{4}{@{\ \,} l }}\hline
Method       & \quad Accuracy & \qquad Dice$_\text{scar}$ & \qquad $G$Dice\\
\hline
LA$_\text{M}$+Otsu~\cite{journal/tmi/ravanelli2013} &$ 0.750 \pm 0.219 $& \quad $ 0.420 \pm 0.106 $ & \quad $ 0.750 \pm 0.188 $\\
LA$_\text{M}$+MGMM~\cite{journal/TBME/liu2017}      &$ 0.717 \pm 0.250 $& \quad $ 0.499 \pm 0.148 $ & \quad $ 0.725 \pm 0.239 $\\
LA$_\text{M}$+LearnGC~\cite{journal/MedIA/li2020}   &$ 0.868 \pm 0.024 $& \quad $ 0.481 \pm 0.151 $ & \quad $ 0.856 \pm 0.029 $\\ 
\hline
LA$_{\text{U\mbox{-}Net}}$+Otsu  &$ 0.604 \pm 0.339 $& \quad $ 0.359 \pm 0.106 $ & \quad $ 0.567 \pm 0.359 $ \\
LA$_{\text{U\mbox{-}Net}}$+MGMM  &$ 0.579 \pm 0.334 $& \quad $ 0.430 \pm 0.174 $ & \quad $ 0.556 \pm 0.370 $ \\
\hline
U-Net$_\text{scar}$-BCE          &$ 0.866 \pm 0.032 $& \quad $ 0.357 \pm 0.199 $ & \quad $ 0.843 \pm 0.043 $\\
U-Net$_\text{scar}$-Dice         &$ 0.881 \pm 0.030 $& \quad $ 0.374 \pm 0.156 $ & \quad $ 0.854 \pm 0.041 $\\
U-Net$_\text{scar}$-SE           &$ 0.868 \pm 0.026 $& \quad $ 0.485 \pm 0.129 $ & \quad $ 0.863 \pm 0.026 $\\
\hline\hline
AJSQnet-BCE                        &$ 0.887 \pm 0.023 $& \quad $ 0.484 \pm 0.099 $ & \quad $ 0.872 \pm 0.024 $\\
AJSQnet-SE                         &$ 0.882 \pm 0.026 $& \quad $ 0.518 \pm 0.110 $ & \quad $ 0.871 \pm 0.024 $\\
AJSQnet-SESA                       &$ 0.867 \pm 0.032 $& \quad $ 0.543 \pm 0.097 $ & \quad $ 0.868 \pm 0.028 $\\
\hline\hline
Inter-Ob                           &$ 0.891 \pm 0.017 $& \quad $ 0.580 \pm 0.110 $ & \quad $ 0.888 \pm 0.022 $\\
\hline
\end{tabular} }\\
\end{table*}

\subsection{Conclusion and future work}
In this sub-chapter, we present two approaches for scar quantification from LGE MRI.
The first approach combines graph-cuts and MS-CNN, referred to as LearnGC, which integrates the multi-scale information of scars and ensures a smooth segmentation results.
More importantly, LearnGC is trying to ignore the wall thickness and project the extracted scars onto the LA surface.
Therefore, it converted the challenging volume-based scar segmentation problem into the relatively easy surface-based scar quantification problem.
However, the pixel-wise quantification on the surface only includes limited information, and tends to be effected by the misalignment between the predicted endocardial surface and the corresponding ground truth.
Therefore, the proposed random shift scheme and MS-CNN are effective to improve the robustness of the proposed method against the LA segmentation errors.
A major limitation of the LearnGC method is the lack of an end-to-end training scheme, i.e., MS-CNN and graph-cuts were separately achieved.
We therefore proposed another approach, i.e., AtrialJSQnet, which can simultaneously achieved LA segmentation and scar quantification.
To eliminate the effect of inaccurate LA segmentation, we learn the spatial information of each pixel on the surface via a newly designed SE loss.
The SE loss and joint optimization were both proved to be effective by observing our experimental results.

A limitation of the two works is that the gold standard was constructed from the manual delineation of only one expert.
Besides, the subjects included in this study are only post-ablation AF patients. 
In future work, we will combine multiple experts to construct the gold standard, and consider both pre- and post-ablation data.
Moreover, we will collect multi-center LGE MRI to explore the generalization ability of LGE MRI segmentation and quantification models.

%%%%%%%%%%%%%%%%%%%%%%%%%%%%%%%%%%%%% DA %%%%%%%%%%%%%%%%%%

\section{Domain adaptation for cross-modality cardiac image segmentation}\label{DA}

% \Leicolor{}{Lei: Hi fuping, after proof read for your sub-chapter, I mainly made following modification. 
% 1. I have modified your description on the cited images, as "is cited from" seems not common usage for this.
% 2. For consistence, I have modified "FIGURE x" into "Figure x".
% 3. "BSSFP" has been changed into "bSSFP".
% ...
% There are some small typos I did not listed here but also modified.}
% \textcolor{blue}{thanks, lei, I have revised them as you commented.}

\subsection{Introduction}
The capacity of model generalization is essential for the application of computer-aided-diagnosis (CAD) system on cardiac image analysis.
In practice, a cardiac segmentation model trained on a specific modality could perform poorly on images from other modalities \cite{dou2018pnp}.
The reason is that there exists nonnegligible gap between the distributions of test and training data, which is known as domain shift \cite{shimodaira2000improving}.
How to transfer the learned anatomical knowledge from one domain to others without labeling new data is an interesting and open problem.
An important research direction is domain adaptation, which aims to reduce the domain discrepancy between the labeled source and unlabeled target data \cite{Csurka2017}.

To date, many domain adaptation approaches have been proposed for cardiac image segmentation. 
Most of them learned modality-invariant features via adversarial training \cite{Chartsias2019Disentangle, dou20173d}.
For example, Dou et al.\cite{dou2018pnp} designed a dilated fully convolutional network (denoted as PnP-AdaNet), which consists of a plug-and-play domain adaptation module to map two domains into a common space. 
It adopted the training scheme of domain adversarial neural networks (DANN) \cite{ganin2016domain}.
The method was validated on MRI-CT cross-modality cardiac segmentation for 2D images.
While PnP-AdaNet extracted domain-invariant latent features from middle layers and achieved promising results, Chen et al. \cite{chen2019synergistic} proposed to implement domain adaptation on both feature and image levels.
Their method, referred to as SIFA, translated images between two domains, and used the cycle-consistency loss for model constrain.
This work was validated to outperform peer methods on cardiac cross-modality segmentation.
In addition,  Ouyang et al.\cite{ouyang2019data} introduced a VAE-based feature prior matching to further adapt their features, and proposed a data efficient method for multi-domain medical image segmentation.

Although adversarial training has shown great potential in domain adaptation, especially for image translation, there still exist drawbacks which degrade its effectiveness and efficiency.
First, the extracted domain-invariant features (DIFs) may not be pure. 
They could contain specific domain information and lead to biased results \cite{cai2019learning}.
For image translation, no attention has been paied on domain-specific features (DSFs), which might be useful to improve the quality of the reconstructed images.
Second, adversarial training reduces the discrepancy implicitly.
It suffers from problems originated from the generative adversarial network (GAN) \cite{goodfellow2014generative}, such as the extra discriminators, complex training process and difficulty of obtaining the Nash equilibrium point \cite{heusel2017gans}.
While many explicit measurements for distribution discrepancy have been designed, such as the Maximum Mean Discrepancy (MMD) \cite{tzeng2014deep} and the moment distance \cite{sun2016deep}, all of them were proposed for classification tasks, and no work has been validated for segmentation, to the best of our knowledge.
What kind of explicit metric is efficient for cross-modality segmentation remains an important and open problem.

\begin{figure}[!t]
	\centering
	\subfloat[The framework of previous researches of domain adaptation in a latent feature space. ]{\includegraphics[width=4.3in]{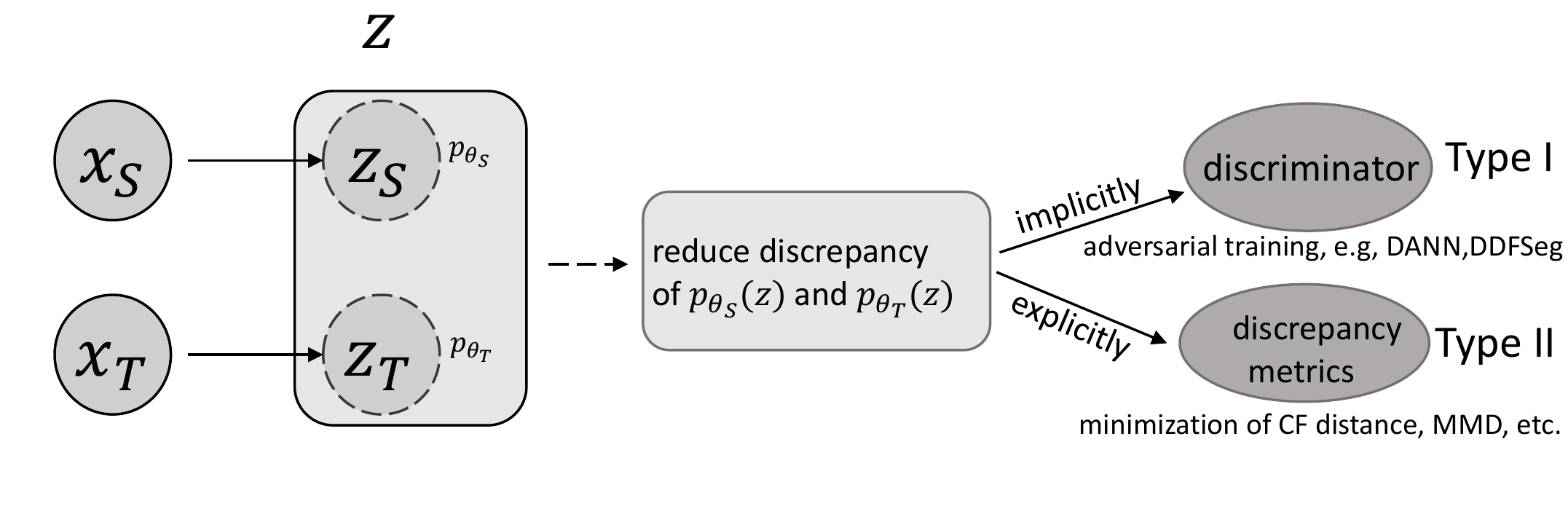}
		\label{fig:previous idea}}
	\hspace{0.055in}
	\subfloat[The idea of the proposed variational approximation method for domain adaptation. ]{\includegraphics[width=4.3in]{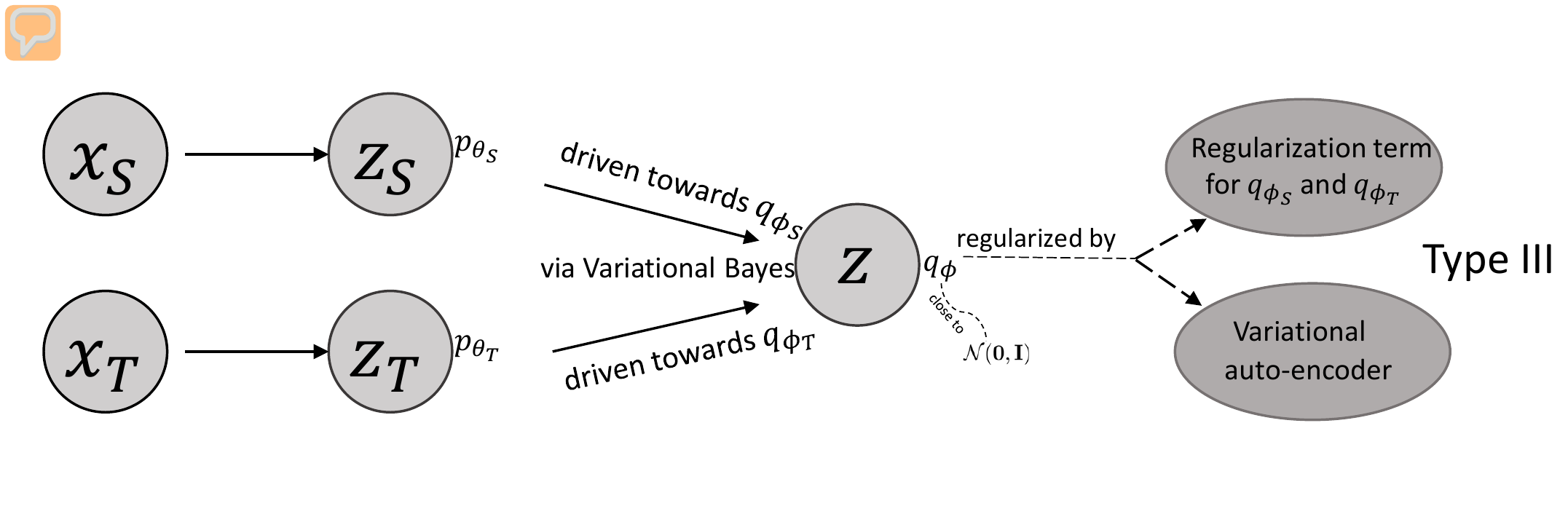}
		\label{fig:this work idea}}
	\caption{Illustration of the difference among three types of methods for domain adaptation. The proposed three frameworks refer to the three types, respectively. Images adopted from Wu et al. \cite{Wu2021TMI}}
	\label{fig:difference}
\end{figure}

To tackle these issues, we proposed three domain adaptation frameworks for cardiac segmentation, which are termed as DDFSeg \cite{PEI2021102078}, CFDNet \cite{Wu2020TMI} and VarDA \cite{Wu2021TMI}, respectively.
Specifically, we first studied feature disentanglement for domain adaptation, and constrained DIFs and DSFs by introducing self-attention and zero-loss.
It uses adversarial training for model optimization, thus can be categorized into Type I, as Figure \ref{fig:difference} (a) illustrated.
Next, we studied the effectiveness of explicit discrepancy metrics for domain adaptation.
A new metric based on the distance between characteristic functions is proposed, and it was validated to be effective in cardiac segmentation tasks.
This metric is denoted as CF distance, and its minimization leads to the reduction of domain discrepancy and the extraction of domain-invariant features.
This method avoids adversarial training, and has a simpler training process and faster model convergence.
As Figure \ref{fig:difference} (a) illustrated, it can be classified as type II approaches.
For both type I and II methods, the domains were mapped into a common latent feature variable $z$. 
The domain discrepancy was then either reduced by adversarial training or explicitly minimization of discrepancy metrics.
While type II methods were validated to be useful for cross-modality cardiac segmentation, especially using the proposed CFDNet, we found that the computation of these metrics are complex.
In practice, we calculate these metrics with marginal distributions instead of joint ones. 
This substitution weakens the constraint for domain-invariant features.
Based on this consideration, we further proposed another type of methods.
As illustrated in Figure \ref{fig:difference} (b), method of Type III drives two domains towards a common parameterized distribution, i.e., $q_{\phi}(z)$, in a latent feature $z$.
As $q_{\phi}$ can be set to independent among its element variables, such as Gaussian distributions, the effect of the aforementioned substitution for metric calculation could be alleviated significantly.
We achieve this approximation using variational auto-encoders, and thus denote the proposed framework as VarDA.
In the following, we will describe the three methods in detail, and present their performances on the cardiac segmentation.

\subsection{Method}

Let $X_S=\{x_S^i\}_{i=1}^{N_{S}}$ be the set of labeled source data, which are independent and identically distributed ($i.i.d.$) samples collected from source domain $x_S$ with distribution $p_{\theta_S}(x)$, and $\theta_S$ is model parameter.
$Y_S=\{y_S^i\}_{i=1}^{N_S}$ denotes the corresponding label of $X_S$.
With $X_S$ and $Y_S$, we can learn a segmentation model for this source domain.
We further collect target data samples from target domain $x_T$, denoted by $X_T=\{x_T^i\}_{i=1}^{N_{T}}$, from a different distributions $p_{\theta_T}(x)$.
The goal of domain adaptation is to transfer the knowledge from the source data, and train a segmentation model for the target data.

\subsubsection{DDFSeg: Disentangle Domain Features for Domain Adaptation and Segmentation}

\begin{figure}[ht]
	\centering
	\includegraphics[width=4.3in]{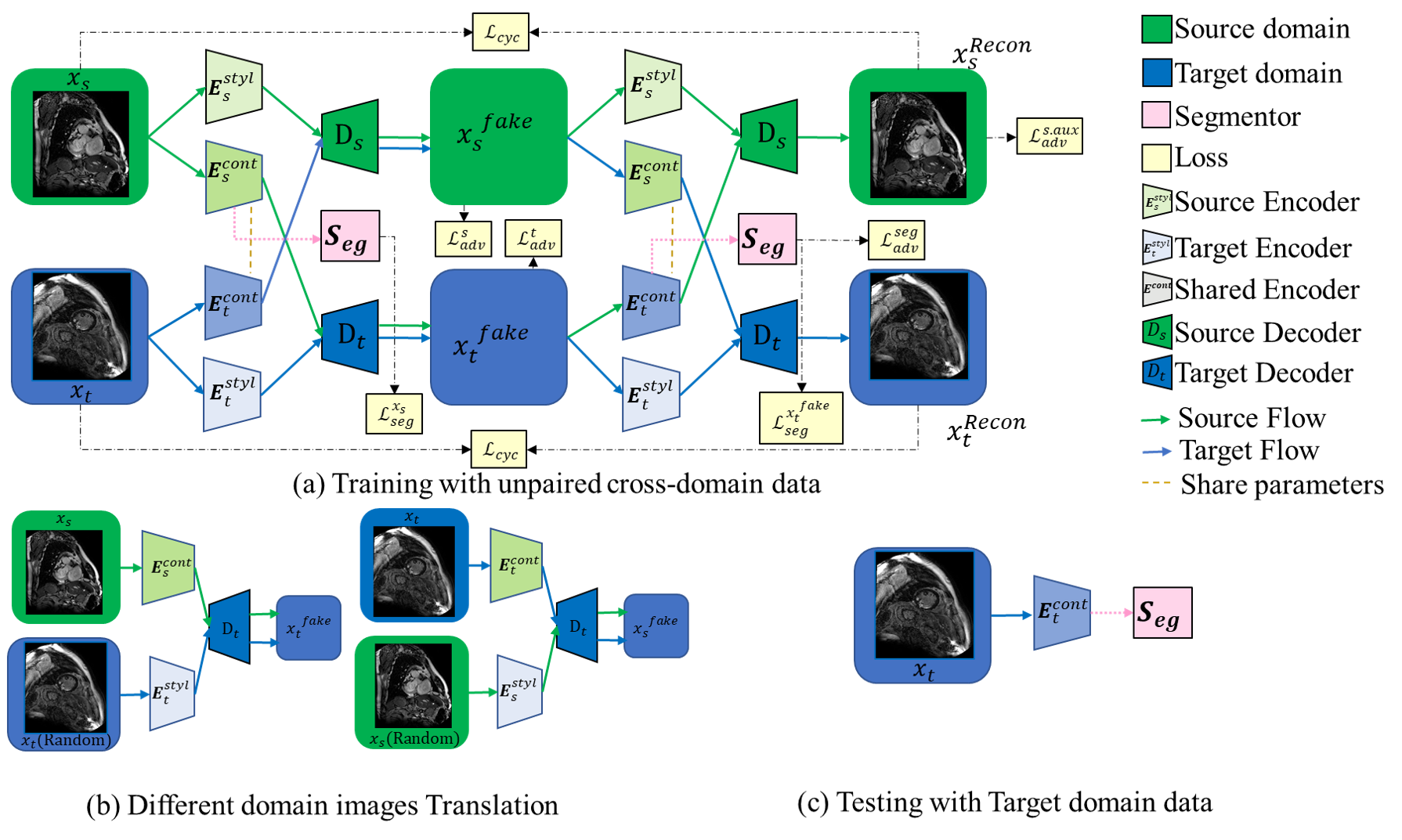}
	\caption{(a) Overview of the framework. (b) Illustration of different domain images translation. (c) Illustration of the segment with target domain image at test time. Images adopted from Pei et al. \cite{PEI2021102078}
	}
	\label{fig:DDFSeg}
\end{figure}

We first study domain adaptation via image translation.
Taking DSFs into account, we propose a new framework.
As Figure \ref{fig:DDFSeg} illustrated, we use four encoders, i.e., $E^{styl}_{S}$ and $E^{cont}_{S}$ for source domain, $E^{styl}_{T}$ and $E^{cont}_{T}$ for target domain, to disentangle each domain into their DIFs and DSFs.
These features are denoted as $z^{c}_{d}=E^{c}_{d}(x_d)$, where $c\in \{styl,cont\}$ and $d\in \{S,T\}$.
These features are then swapped and decoded into images with the anatomical structures maintained and the style exchanged, using two decoders $D_S$ and $D_T$.
Mathematically, the new generated images can be expressed as $x_S^{fake}=D_S(z^{styl}_{S},z^{cont}_{T})$, and $x_T^{fake}=D_T(z^{styl}_{T},z^{cont}_{S})$.
$x_S^{fake}$ and $x_T^{fake}$ are further encoded and decoded again for image reconstruction, which can be seen as a modified version of CycleGAN \cite{zhu:unpaired}.
The reconstructed images are denoted as $x_s^{recon}$ and $x_t^{recon}$, with $x_S^{recon}=D_S(E_S^{styl}(x_S^{fake}),E_T^{cont}(x_T^{fake}))$ and $x_T^{recon}=D_T(E_T^{styl}(x_T^{fake}),E_S^{cont}(x_S^{fake}))$.
To enhance the DIFs and DSFs, we introduce the techniques of self-attention and zero-loss .
In addition, extra discriminator is used to constrain the anatomical shape of segmentation output.
Hence, the total loss function consists of three parts, i.e., image translation loss, zero-loss, and segmentation loss.

For image translation, we adopt cycle consistency loss to achieve image reconstruction, and use several discriminators to force the generated images to be as real as possible.
The cycle consistency loss is defined as follows,

\begin{equation}
\begin{aligned}
&\mathcal{L}_{cyc}\left(E_S^{cont}, E_S^{styl}, E_T^{cont}, E_T^{styl}, D_S, D_T\right)\\=&
\mathbb{E}_{x_{S},x_{T} \sim P\left(X_{S},X_{T}\right)}\left\|x_S^{recon}-x_{S}\right\|_{1}+\mathbb{E}_{x_{S},x_{T} \sim
	P\left(X_{S},X_{T}\right)}\left\|x_T^{recon}-x_{T}\right\|_{1}.
\end{aligned}
\end{equation}

To force $x_S^{fake}$ and $x_T^{fake}$ to be real, we introduce two discriminators $Dis_S$ and $Dis_T$ for source and target domains, respectively.
The objective functions for adversarial training can be formulated as follows,

\begin{equation}
\begin{aligned}
\min _{\left(E_S^{cont}, E_T^{styl}, D_T\right)} \max _{Dis_T}
\mathcal{L}_{adv}^{T}=&
\mathbb{E}_{x_{T} \sim P\left(X_{T}\right)}\left[\log Dis_T\left(x_{T}\right)\right]+\\
& \mathbb{E}_{x_{S},x_{T} \sim P\left(X_{S},X_{T}\right)}\left[\log\left(1-Dis_T\left(x_T^{fake}\right)\right)\right],
\end{aligned}
\end{equation}

\begin{equation}
\begin{aligned}
\min _{\left(E_T^{cont}, E_S^{styl}, D_S\right)} \max _{Dis_S}
\mathcal{L}_{adv}^{S}=&
\mathbb{E}_{x_{S} \sim P\left(X_{S}\right)}\left[\log Dis_S\left(x_{S}\right)\right]+\\
& \mathbb{E}_{x_{S},x_{T} \sim P\left(X_{S},X_{T}\right)}\left[\log\left(1-Dis_S\left(x_S^{fake}\right)\right)\right].
\end{aligned}
\end{equation}

Moreover, to further enhance the extracted DIFs to be domain-invariant, we add an auxiliary task to the source discriminator $Dis_S$ to differentiate $x_s^{fake}$ and $x_s^{recon}$. 
The objective function is defined as follows,
\begin{equation}
\begin{aligned}
\min _{E_T^{cont}} \max _{Dis_S}
\mathcal{L}_{adv}^{S.aux}=&
\mathbb{E}_{x_S^{recon} \sim P\left(x_S^{recon}\right)}\left[\log Dis_S\left(x_S^{recon}\right)\right]+\\
&\mathbb{E}_{x_S^{fake} \sim P\left(x_S^{fake}\right)}
\left[\log \left(1-Dis_S\left(x_S^{fake}\right)\right)\right].
\end{aligned}
\end{equation}

The zero-loss is used to to constrain the encoders $E_S^{styl}$ and $E_T^{styl}$ to force the extracted information from target and source images to be zero.
Hence, the losses can be formulated as follows,
\begin{equation}
\begin{aligned}
\mathcal{L}_{\text {zero}}^{T}\left(E_T^{styl}\right)=\mathbb{E}_{x_S\sim P\left(X_S\right)}\left[\left\|E_T^{styl}\left(x_{S}\right)\right\|_{1}\right] .
\end{aligned}\label{S7.2:eq.5}
\end{equation}
and 
\begin{equation}
\begin{aligned}
\mathcal{L}_{\text {zero}}^{S}\left(E_S^{styl}\right)=\mathbb{E}_{x_T\sim P\left(X_T\right)}\left[\left\|E_S^{styl}\left(x_{T}\right)\right\|_{1}\right] .
\end{aligned}\label{S7.2:eq.6}
\end{equation}
Combining \eqref{S7.2:eq.5} \eqref{S7.2:eq.6}, we have the total zero-loss,
\begin{equation}
\begin{aligned}
\mathcal{L}_{\text {zero}}=\mathcal{L}_{\text {zero}}^{T}+\mathcal{L}_{\text {zero}}^{S}.
\end{aligned}
\end{equation}

The third types of losses is designed for semantic segmentation.
The segmentation module predicts the labels from the latent features $z^{cont}_{S}$ and $z^{cont}_{T}$.
The first segmentation loss can be formulated as follows,
	\begin{equation}
	\begin{aligned}
	\mathcal{L}_{seg}^{x_{S}}\left(E_S^{cont},\mathcal{S}_{seg}\right)=\mathbb{E}_{x_{S},y_{S}\sim P\left(X_S,Y_{S}\right)}\left[\mathcal{C}\left(y_{S},\widehat{y}_{S}\right)+\alpha\cdot Dice\left(y_{S},\widehat{y}_{S}\right)\right],
	\end{aligned}
	\end{equation} 
	where$\widehat{y}_{S}$ is the prediction of $x_{S}$, $\widehat{y}_{S}=\mathcal{S}_{seg}(z_s^{cont})$, $\mathcal{C}\left(y_{S},\widehat{y}_{S}\right)$ is the cross-entropy loss, $Dice\left(y_{S},\widehat{y}_{S}\right)$ the Dice loss, and $\alpha$ is the hyper-parameter.
	
We further use the generated images $x_{t}^{fake}$, which contains the same anatomical information as $x_{S}$, to train the segmentation module.
The loss is then formulated as follows,
\small{
\begin{equation}
    \begin{aligned}
    \mathcal{L}_{seg}^{x_{T}^{fake}}\left(E_T^{cont},\mathcal{S}_{seg}\right)=\mathbb{E}_{x_{T}^{fake},y_{S}\sim P\left(X_{T}^{fake},Y_S\right)}\left[ \mathcal{C}\left(y_{S},\widehat{y}_{S}^{fake}\right)+\alpha\cdot Dice\left(y_{S},\widehat{y}_{S}^{fake}\right)\right],
    \end{aligned}
\end{equation}}
where $\widehat{y}_{S}^{fake}$ is the prediction of $x_{T}^{fake}$.
	
We have the total segmentation loss as follows,
\begin{equation}
    \begin{aligned}
    \mathcal{L}_{seg}=\mathcal{L}_{seg}^{x_{S}}+\mathcal{L}_{seg}^{x_{T}^{fake}}.
    \end{aligned}
\end{equation}
	
In addition, we introduce another discriminator $Dis^{seg}$ to constrain the shape of segmentation output of target images to be similar to that of source images.
The objective function is defined as follows,
\begin{equation}
    \begin{aligned}
    \min _{\left(E_T^{cont}, \mathcal{S}_{seg}\right)} \max _{Dis^{seg}}
    \mathcal{L}_{adv}^{seg}=
    &\mathbb{E}_{x_{T}^{fake} \sim P\left(x_{T}^{fake}\right)}\left[\log Dis^{seg}\left(\mathcal{S}_{seg}\left(E_T^{cont}\left(x_{T}^{fake}\right)\right)\right)\right]+\\
    &\mathbb{E}_{x_{T} \sim P\left(X_{T}\right)}\left[\log\left(1- Dis^{seg}\left(\mathcal{S}_{seg}\left(E_T^{cont}\left(x_{T}\right)\right)\right)\right)\right].
    \end{aligned}
\end{equation}

Combining all the aforementioned losses, we have the total loss as follows,
\begin{equation}
    \begin{aligned}
    \mathcal{L}=&\lambda_{1}\mathcal{L}_{adv}^{t}\left(E_s^{cont}, E_t^{styl}, D_t, Dis_t\right)+\\
    &\lambda_{2}\mathcal{L}_{adv}^{s}\left(E_t^{cont}, E_s^{styl}, D_s, Dis_s\right)+\\
    &\lambda_{3}\mathcal{L}_{cyc}\left(E_s^{cont}, E_s^{styl}, E_t^{cont}, E_t^{styl}, D_s, D_t\right)+\\
    &\lambda_{4}\mathcal{L}_{adv}^{s.aux}\left(E_t^{cont}, Dis_s\right)+\lambda_{5}\mathcal{L}_{\text {zero}}\left(E_t^{styl},E_s^{styl}\right)+\\
    &\lambda_{6}\mathcal{L}_{seg}\left(E_s^{cont}, E_t^{cont}, S\right)+\lambda_{7}\mathcal{L}_{adv}^{seg}\left(E_t^{cont}, S, Dis^{seg}\right).
    \end{aligned}
\end{equation}
In experiments, we set $\lambda_{adv}^{1}=1.0$, $\lambda_{adv}^{2}=1.0$, $\lambda_{3}=1.0$, $\lambda_{4}=0.1$, $\lambda_{5}=0.01$, $\lambda_{6}=0.1$, and $\lambda_{7}=0.1$.

\subsubsection{CFDNet: Characteristic Function Distance for Unsupervised Domain Adaptation}

\begin{figure}[!t]
	\centering
	\framebox{\includegraphics[width=1.0\textwidth,height=0.5\textwidth]{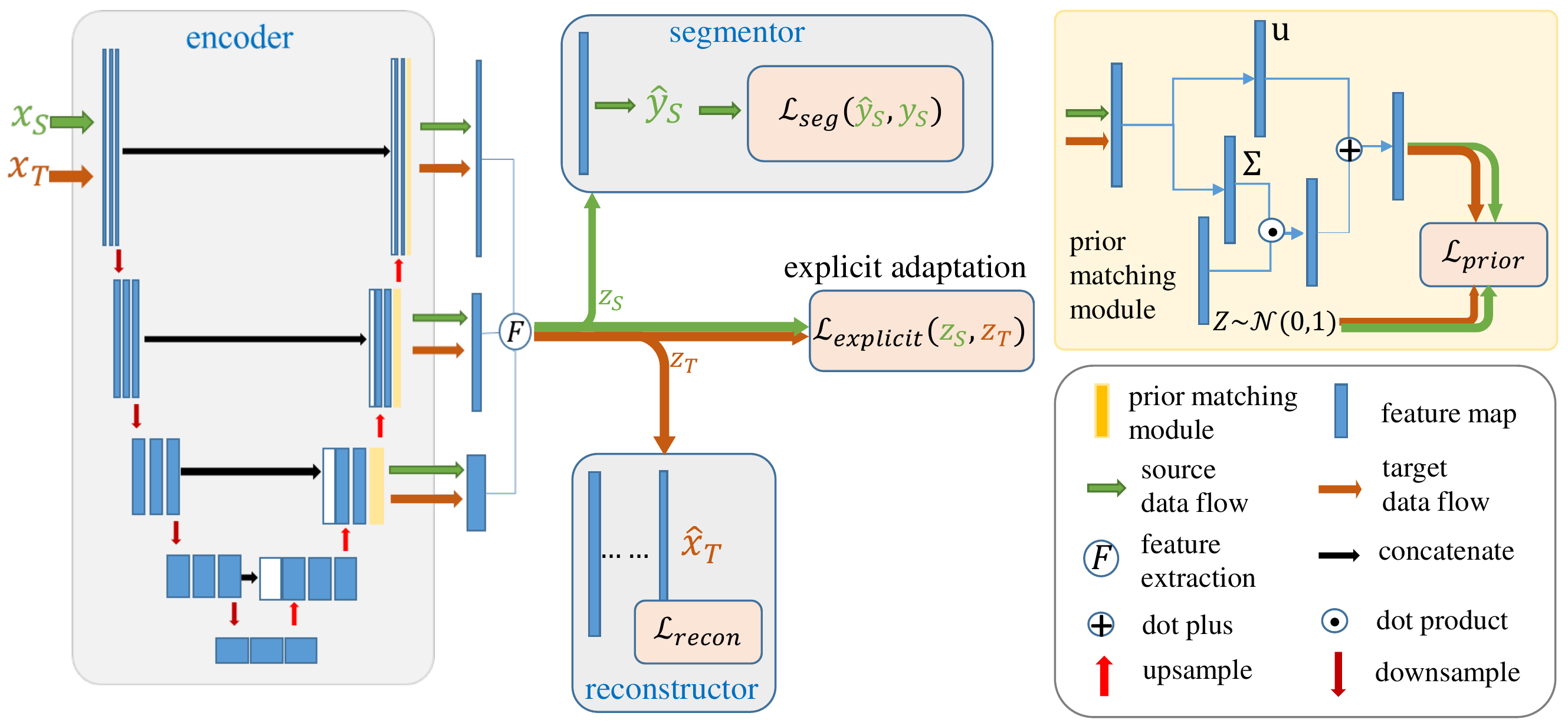}}\\[0ex]
	\caption{Framework of the proposed domain adaptation method for medical image segmentation. Images adopted from Wu et al. \cite{Wu2020TMI}
%The readers are referred to Section~\ref{adversarial} for details.
	}
	\label{fig:frameworkcfd}
\end{figure}

Beside the adversarial training, we further study the effectiveness of explicit metric for domain adaptation.
We propose a new metric which measures the distance between the characteristic functions of the latent features from source and target domains.
Based on this CF distance, we propose a framework for cardiac segmentation, denoted as CFDNet.
Figure \ref{fig:frameworkcfd} illustrated the whole framework. The encoder extracts the latent features $z_S\in \mathbb{R}^n$ and $z_T\in \mathbb{R}^n$ respectively from the source and target data. The segmentor module outputs the segmentation results from the latent features. The reconstructor module reconstructs the target images. The prior matching module regularizes the prior distributions of $z_S$ and $z_T$ to be close to $\mathcal{N}(0,I)$. The explicit adaptation module computes the domain discrepancy explicitly.
Next we describe each module in detail.

Calculating the discrepancy between the distributions of two domains in the latent feature space, i.e., $p_{z_S}(z)$ and $p_{z_T}(z)$, is difficult, because they are unknown. We instead  estimate the distance of their CFs, $f_{z_S}(\vec{t})$ and $f_{z_T}(\vec{t})$.
The loss function is given by,
\begin{equation}
\mathcal{D}_{CF}(z_S,z_T;U)= \int_{-U}^{U}||f_{z_S}(\vec{t})-f_{z_T}(\vec{t})||^2 dt.
\end{equation}
where $U\in \mathbb{R_+}^n$.
 
In practice, we solely compute the loss for a single point $U_a=[a,\cdots,a]\in \mathbb{R_+}^n$ using the mini-batch of samples.
The CF distance is then estimated by,
\begin{flalign}\label{3.4}	
    \mathcal{D}_{CF}(z_S,z_T;U_a)\approx& \frac{1}{M^2}\sum_{p=1}^{M}\sum_{q=1}^{M}k(z_S^{p},z_S^{q})+\notag \\ &\frac{1}{M^2}\sum_{p=1}^{M}\sum_{q=1}^{M}k(z_T^{p},z_T^{q})-\frac{2}{M^2}\sum_{p=1}^{M}\sum_{q=1}^{M}k(z_S^{p},z_T^{q}),
\end{flalign}
where $k(z_S,z_T)$=$\prod_{k=1}^{n}\!\!\frac{2\sin[(z_{S_k}-z_{T_k})a]}{z_{S_k}-z_{T_k}}$, for $\forall z_S,z_T \in \mathbb{R}^n$; $z_{S_k}$ is the $k$-th element of $z_{S}$, and likewise for $z_{T_k}$;
$M$ is the number of samples.

To simplify the calculation, we adopt a sliced version of CFD distance as a substitution, which is defined as follows,
\begin{flalign}\label{3.7}
    \mathcal{L}_{SCF}(z_S,z_T;U)\approx\frac{1}{n}\sum_{i=1}^{n}\mathcal{D}_{CF}(z_{S_i},z_{T_i};U).
\end{flalign}

We further introduce the mean value matching to enforce domain adaptation.
The mean loss is defined as follows,
\begin{flalign}\label{3.8}
    \mathcal{L}_{mean}=\parallel \mathbb{E}_{p_{z_S}(z)}(z)-\mathbb{E}_{p_{z_S}(z)}(z)\parallel^2.
\end{flalign}

Combining the sliced CF distance with the mean loss, we have the explicit domain discrepancy loss,
\begin{flalign}\label{eq:cfdloss}
    \mathcal{L}_{explicit}=\mathcal{L}_{CFD}=\beta_1 \mathcal{L}_{SCF}+\beta_2 \mathcal{L}_{mean},
\end{flalign}
where $\beta_1$ and $\beta_2$ are hyperparameters.

Moreover, we introduce the technique of prior matching.
We use a variational auto-encoder to map two domains into a common latent space $z$, with posterior distributions being subject to $q_{\phi_{S/T}}(z|x)=\mathcal{N}(u_{S/T},\Sigma_{S/T})$, where $u_{S/T}=(u_{S/T}^{1},\cdots, u_{S/T}^{n})\in \mathbb{R}^n$, $\Sigma_{S/T}={\rm diag}(\lambda_{S/T}^{1},$ $\cdots,\lambda_{S/T}^{n})\in \mathbb{R}^{n\times n}$.
The prior matching loss is fornulated as follows,
\begin{flalign}
    \mathcal{L}_{prior}=E_{ p_{x_S}(x)}[KL(q_{\phi_S}(z|x)||\mathcal{N}(\bm{0},I))]+
    E_{ p_{x_T}(x)}[KL(q_{\phi_T}(z|x)||\mathcal{N}(\bm{0},I))],
\end{flalign}
where $KL(\cdot||\cdot)$ denotes the Kullback-Leibler ($KL$) divergence, and $\bm{0}\in \mathbb{R}^{n}$ is a zero vector.

To constrain the feature of target images, we add a reconstruction loss, denoted as $\mathcal{L}_{recon}$.
We use the cross entropy loss for the segmentation loss from the labeled source domain, and denote it as  $\mathcal{L}_{seg}(\widehat{y}_S,y_S)$.
Then, the total loss of the proposed CFDNet is formulated as follows,
\begin{equation}\label{3.1}
    \mathcal{L} = \alpha_1 \mathcal{L}_{seg}+\alpha_2 \mathcal{L}_{prior}+\alpha_3 \mathcal{L}_{recon}+\alpha_4 \mathcal{L}_{explicit},
\end{equation}
where $\alpha_1$, $\alpha_2$, $\alpha_3$ and $\alpha_4$ are the parameters.

\subsubsection{VarDA: Domain Adaptation via Variational Approximation}

Although the proposed CF distance is validated to be effective for domain adaptation, the substitution of its calculation using marginal distributions leads to a weaker constraint on the features.
Moreover, the prior matching technique is not very useful as expected, due to the two-step sampling for estimation of the CF distance.
Based on these observation, we further proposed another domain adaptation framework, which drives two domains towards a common parameterized distribution via variational approximation.

\begin{figure}[t]
	\centering
	\includegraphics[width=1.0\textwidth]{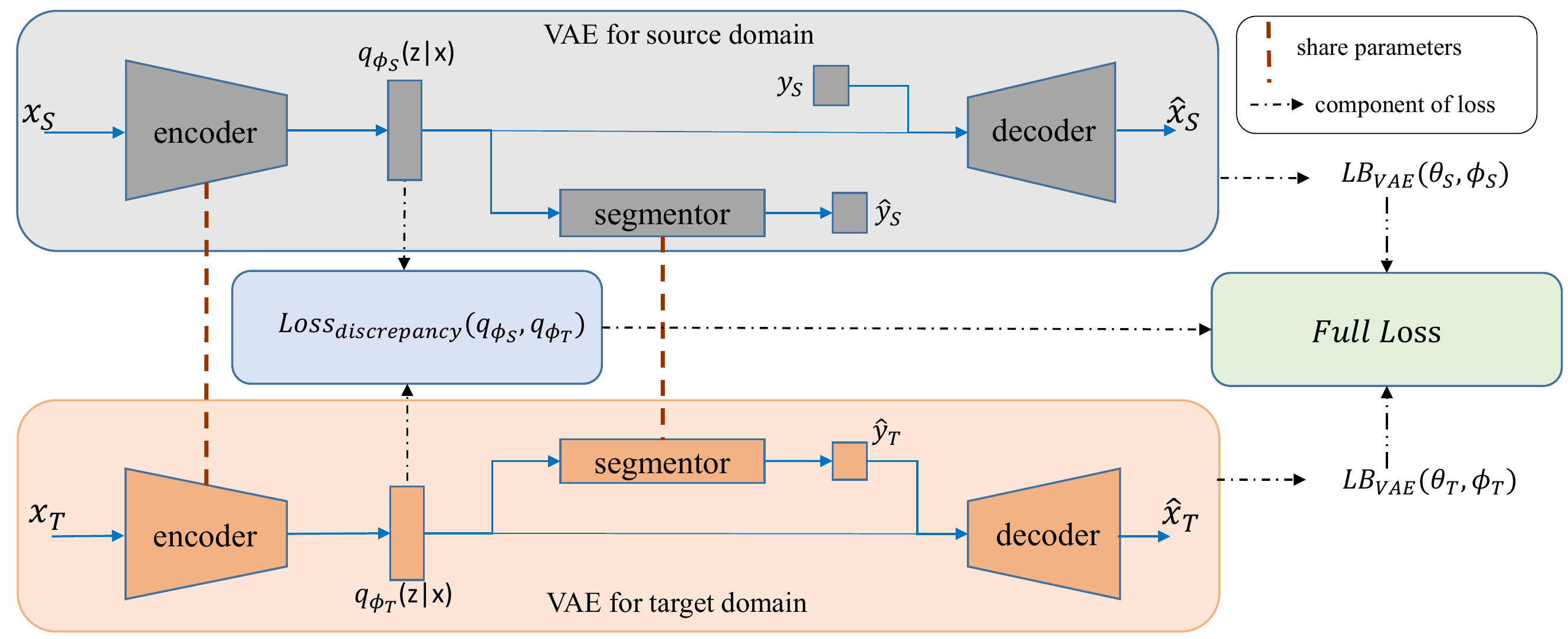}%\\[-1ex]
	\caption{Framework of the proposed VarDA. Image adopted from Wu et al. \cite{Wu2021TMI}
	}
	\label{fig:frameworkvarda}
\end{figure}

As Figure \ref{fig:frameworkvarda} illustrated, the proposed VarDA framework consists of three modules, i.e., two VAE for each domains, and a module to compute the regularization term on domain discrepancy.
The objective functions of the two VAEs are denoted as $LB_{VAE}(\theta_S,\phi_S)$ and $LB_{VAE}(\theta_T,\phi_T)$, respectively.
The regularization term for $q_{\phi_S}(z)$ and $q_{\phi_T}(z)$ is denoted as $Loss_{discrepancy}(q_{\phi_S},q_{\phi_T})$.
The total loss function of VarDA is then formulated by,
\begin{equation}\label{2.1}\begin{array}{l@{\ }l}
Full\,Loss(\omega)=&-\alpha_1 LB_{VAE}(\theta_S,\phi_S) -\alpha_2 LB_{VAE}(\theta_T,\phi_T)\\
&+\alpha_3 Loss_{discrepancy}(q_{\phi_S},q_{\phi_T}),
\end{array}\end{equation}
where $\omega=(\theta_S,\phi_S,\theta_T,\phi_T)$ are the parameters to be optimized, and $\alpha_1$, $\alpha_2$, $\alpha_3$ are the trade-off parameters.

Similar to the original VAE, we have the following objective function for the source domain,
\begin{align}
    &LB_{VAE}(\theta_S,\phi_S)=-D_{KL}(q_{\phi_S}(z|x)\parallel p_{\theta_S}(z))\quad \notag \\
    & +E_{\log q_{\phi_S}(z|x)}[ p_{\theta_S}(x|y,z)]+ E_{q_{\phi_S}(z|x)}[\log p_{\theta_S}(y|z)],
    \label{2.8}
\end{align}
where $D_{KL}(q || p)$ is the $KL$ divergence of $q$ and $p$.
The second term $E_{q_{\phi_S}(z|x)}[\log$ $ p_{\theta_S}(x|y,z)]$ can be modeled by the image reconstruction.
The third term $E_{q_{\phi_S}(z|x)}[\log $ $p_{\theta_S}(y|z)]$ is modeled by the segmentor.

We estimate $LB_{VAE}(\theta_S,\phi_S)$ using mini-batch samples as follows,
\begin{align}
    \widetilde{\mathcal{L}}_S(\theta_S,&\phi_S;x^i,y^i)=-D_{KL}(q_{\phi_S}(z^i|x^i)\parallel p_{\theta_S}(z^i))\notag \\
    &+\frac{1}{L}\sum_{l=1}^{L}\big[\log p_{\theta_S}(x^i|y^i,z^{(i,l)})+\log p_{\theta_S}(y^i|z^{(i,l)})\big],
    \label{2.12}
\end{align}
where $z^{(i,l)}=g_{\phi_S}(\epsilon^{(i,l)},x^i)$, with $\epsilon^{(i,l)}\sim p(\epsilon)$, $L$ is the number of samples, and $z=g_{\phi_S}(\epsilon,x)$ is a differentiable transformation with $\epsilon\sim p(\epsilon)$.

Similarly, we have the variational lower bound for the target domain as follows,
\begin{align}
    &LB_{VAE}(\theta_T,\phi_T)=-D_{KL}(q_{\phi_T}(z|x)\parallel p_{\theta_T}(z))\notag \\
    & +E_{q_{\phi_T}(z|x)}[\log p_{\theta_T}(x|\widehat{y},z)]+ E_{q_{\phi_T}(z|x)}[\log p_{\theta_T}(\widehat{y}|z)].
    \label{2.15}
\end{align}

Finally, we force the approximations $q_{\phi_S}(z)$ and $q_{\phi_T}(z)$ to be the same one, and thus have the regularization term using $l_2$ norm of their distance as follows,
\begin{flalign}
    D&(\phi_S,\phi_T)=\int[q_{\phi_S}(z)-q_{\phi_T}(z)]^2{\rm d}z  \notag \\ %~~~~~(by~ \zxhrefeq{2.2})
    &\approx \int\Big[\frac{1}{M}\sum_{i=1}^{M}q_{\phi_S}(z|x_S^i)-\frac{1}{M}\sum_{j=1}^{M}q_{\phi_T}(z|x_T^i)\Big]^2{\rm d}z \notag\\
    %&=\frac{1}{M^2}\sum_{i=1}^{M}\sum_{j=1}^{M}k(x_S^i,x_S^j)+\frac{1}{M^2}\sum_{i=1}^{M}\sum_{j=1}^{M}k(x_T^i,x_T^j)\notag\\
    %&-\frac{2}{M^2}\sum_{i=1}^{M}\sum_{j=1}^{M}k(x_S^i,x_T^j),
    &=\frac{1}{M^2}\sum_{i=1}^{M}\sum_{j=1}^{M}\left[k(x_S^i,x_S^j)+k(x_T^i,x_T^j)-2k(x_S^i,x_T^j)\right],
    \label{2.3}
\end{flalign}
where $k(x_S^i,x_T^j)=\int q_{\phi_S}(z|x_S^i)\cdot q_{\phi_T}(z|x_T^j){\rm d}z$.
Let $q_{\phi_S}(z|x_S^i)$ and $q_{\phi_T}(z|x_T^j)$ subject to $N(u_S^i,\Sigma_S^i)$ or $N(u_T^j,\Sigma_T^j)$, one can obtain that
\begin{equation}
    k(x_S^i,x_T^j)=\frac{e^{-\frac{1}{2}\sum_{l=1}^{n}\frac{(u_{S_l}^i-u_{T_l}^j)^2}{\lambda_{S_l}^i+\lambda_{T_l}^j}}}{(2\pi)^{\frac{n}{2}}\cdot (\prod_{l=1}^{n}(\lambda_{S_l}^i+\lambda_{T_l}^j))^{\frac{1}{2}}},
    \label{2.5}
\end{equation}
where $u_{S_l}^i$ is the $l$-th element of $u_{S}^i$.

As the computation for this regularization term is complex, similar to the sliced CF distance, we use the marginal distributions of $z_S$ and $z_T$ to calculate the distance.
The substitution is as follows,
\begin{align}
    \widetilde{\mathcal{D}}(\phi_S,\phi_T)=\sum_{i=1}^{n}
    \int[q_{\phi_S}(z_i)-q_{\phi_T}(z_i)]^2{\rm d}z.
    \label{eq2.55}
\end{align}

Based on the three loss functions discussed above, we have the total loss of VarDA as follows,
\begin{align}
    \widetilde{\mathcal{H}}(\omega)=
    &-\alpha_1\cdot\widetilde{\mathcal{L}}_S(\theta_S,\phi_S;X_S,Y_S)\notag \\
    &-\alpha_2 \cdot \widetilde{\mathcal{L}}_T(\theta_T,\phi_T;X_T,\widehat{Y}_T) +\alpha_3 \cdot \widetilde{\mathcal{D}}(\phi_S,\phi_T). \label{2.16}
\end{align}

\subsection{Data and results}

In this section, we present performances of the three proposed frameworks on two cardiac segmentation tasks, i.e., CT-MR cross modality cardiac segmentation, and C0-LGE multi-sequence CMR segmentation.

\subsubsection{Data}

We used two datasets from two public challenges, i.e., the CT-MR dataset from MM-WHS challenge \cite{journal/mia/Zhuang2016,ZHUANGEvaluation}, and the the bSSFP and LGE CMR images from MS-CMRSeg challenge $\footnote{http://www.sdspeople.fudan.edu.cn/zhuangxiahai/0/mscmrseg19/}$.
The former dataset was from different subjects, while the later was paired images from the same subjects and we shuffled them to be unpaired.

\textbf{CT-MR dataset}: 
This dataset consists of  52 cardiac CT images and 45 MR images, of which 20 CT images and 20 MR images were from the MM-WHS challenge, and the others were from an open data source \cite{schaap2009standardized}.
For each 3D image, 16 slices from the long-axis view around the center of left ventricular cavity were selected, cropped with size of $192\times192$ pixel around the center of heart. 
All methods were validated on these 2D slices for the segmentation of left ventricular cavity (LV) and left ventricular myocardium (MYO).

\textbf{MS-CMRSeg dataset}:
This dataset consists of 45 paired bSSFP CMR and LGE CMR images, among which 5 LGE CMR images were provided with labels for validation, and the ground truths of other 40 images were not available.
The target is to learn knowledge from the labeled bSSFP CMR, and transfer it to LGE CMR images for the prediction of LV, MYO and RV.

\subsubsection{Comparison study for DDFSeg}
We compared DDFSeg with other four methods: 
(1) \textbf{Unet(supervised):} A U-net was trained with labeled target data  in a supervised manner. 
(2) \textbf{Unet(NoAdapt):} A U-net was trained with the source data, and then applied directly on the target images. 
(3) \textbf{CycleSeg:} We used CycleGAN \cite{zhu:unpaired} for image translation. The generated fake target images were then used for model training, and 
(4) \textbf{SIFA} \cite {chen2019synergistic}.

\begin{table}[tb]
	\centering
	\caption{Comparison results  for DDFSeg on  the CT-MR cardiac dataset in both directions of domain adaptation. This table adopted from Pei et al. \cite{PEI2021102078}}
	\resizebox{\textwidth}{!}{ %
		
		\begin{tabular}{lllllllll}
			\hline
			\multicolumn{9}{c}{CT$\longrightarrow$MR}                                                                                                                                                                                                                                                    \\ \hline
			\multicolumn{1}{l|}{\multirow{2}{*}{Method}} & \multicolumn{2}{c}{MYO}  & \multicolumn{2}{c}{RV}      & \multicolumn{2}{c}{LV}       & \multicolumn{2}{c}{Mean}     \\ \cline{2-9}
			\multicolumn{1}{l|}{}                        & \multicolumn{1}{c}{Dice(\%)} & \multicolumn{1}{c|}{ASD(mm)} & \multicolumn{1}{c}{Dice(\%)} & \multicolumn{1}{c|}{ASD(mm)} & \multicolumn{1}{c}{Dice(\%)} & \multicolumn{1}{c|}{ASD(mm)} & \multicolumn{1}{c}{Dice(\%)} & \multicolumn{1}{c}{ASD(mm)} \\ \hline
			\multicolumn{1}{l|}{Unet(supervised)}  & 77.1$\pm$10.2            & \multicolumn{1}{l|}{8.6$\pm$5.4}  &  86.8$\pm$11.2   &    \multicolumn{1}{l|}{4.9$\pm$3.1}     &  90.3$\pm$8.4                    & \multicolumn{1}{l|}{2.5$\pm$5.4}     &   84.8$\pm$11.4                       &  5.3$\pm$4.5         \\\hline
			\multicolumn{1}{l|}{Unet(NoAdapt)}   &   23.8$\pm$24.1                       & \multicolumn{1}{l|}{17.2$\pm$8.5}     &  64.7$\pm$22.1   & \multicolumn{1}{l|}{12.6$\pm$7.9}     & 72.0$\pm$19.7 & \multicolumn{1}{l|}{8.7$\pm$5.1}     &  53.4$\pm$30.6  &  12.8$\pm$8.1         \\
			\multicolumn{1}{l|}{CycleSeg}  &  53.2$\pm$17.1     & \multicolumn{1}{l|}{11.8$\pm$5.1}     & 79.2$\pm$13.1  & \multicolumn{1}{l|}{8.9$\pm$4.7}     & 81.3$\pm$11.8  & \multicolumn{1}{l|}{6.6$\pm$3.6}     &  71.2$\pm$19.1  &  9.1$\pm$5.0          \\
			\multicolumn{1}{l|}{SIFA}  &  67.3$\pm$11.4          & \multicolumn{1}{l|}{\textbf{8.2$\pm$5.3}}     & \textbf{84.2$\pm$11.5}          & \multicolumn{1}{l|}{5.3$\pm$2.8}     & 87.6$\pm$8.9    & \multicolumn{1}{l|}{4.6$\pm$2.3}     &   79.6$\pm$13.9    &  \textbf{6.0$\pm$4.0}   \\

			\multicolumn{1}{l|}{DDFseg}   &    \textbf{71.3$\pm$10.6}   & \multicolumn{1}{l|}{9.7$\pm$5.7}     & 83.2$\pm$11.7    & \multicolumn{1}{l|}{\textbf{4.6$\pm$2.4}}     & \textbf{87.7$\pm$10.4}             & \multicolumn{1}{l|}{3.8$\pm$1.9}     &  \textbf{80.7$\pm$12.9}   &  6.0$\pm$4.5          \\ \hline
			
			\multicolumn{9}{c}{MR$\longrightarrow$CT}                                                                                                                                                                                                                                                    \\ \hline
			\multicolumn{1}{l|}{\multirow{2}{*}{Method}} & \multicolumn{2}{c}{MYO}  & \multicolumn{2}{c}{RV}      & \multicolumn{2}{c}{LV}       & \multicolumn{2}{c}{Mean}     \\ \cline{2-9}
			\multicolumn{1}{l|}{}                        & \multicolumn{1}{c}{Dice(\%)} & \multicolumn{1}{c|}{ASD(mm)} & \multicolumn{1}{c}{Dice(\%)} & \multicolumn{1}{c|}{ASD(mm)} & \multicolumn{1}{c}{Dice(\%)} & \multicolumn{1}{c|}{ASD(mm)} & \multicolumn{1}{c}{Dice(\%)} & \multicolumn{1}{c}{ASD(mm)} \\ \hline
			\multicolumn{1}{l|}{Unet(supervised)}  & 84.1$\pm$5.0            & \multicolumn{1}{l|}{3.2$\pm$2.4}  &  89.2$\pm$6.7   &    \multicolumn{1}{l|}{3.9$\pm$2.3}     &  90.6$\pm$10.6                    & \multicolumn{1}{l|}{3.9$\pm$3.3}     &   88.0$\pm$8.3                       &  3.6$\pm$2.7         \\\hline
			\multicolumn{1}{l|}{Unet(NoAdapt)}   &   10.6$\pm$9.1                       & \multicolumn{1}{l|}{22.2$\pm$8.0}     &  56.0$\pm$12.4   & \multicolumn{1}{l|}{18.3$\pm$7.0}     & 56.2$\pm$16.7 & \multicolumn{1}{l|}{17.0$\pm$5.0}     &  40.9$\pm$25.1  &  19.2$\pm$7.1        \\
			\multicolumn{1}{l|}{CycleSeg}                       &  51.3$\pm$15.4     & \multicolumn{1}{l|}{6.6$\pm$3.8}     & \textbf{83.3$\pm$7.7}  & \multicolumn{1}{l|}{8.4$\pm$2.9}     & 79.3$\pm$15.3  & \multicolumn{1}{l|}{8.3$\pm$3.9}     &  71.3$\pm$19.5  &  7.8$\pm$3.7                          \\
			\multicolumn{1}{l|}{SIFA}  &  56.6$\pm$12.4  & \multicolumn{1}{l|}{\textbf{6.8$\pm$3.8}}     &  80.0$\pm$8.3 & \multicolumn{1}{l|}{8.0$\pm$2.7}     & 82.6$\pm$12.6  & \multicolumn{1}{l|}{\textbf{7.8$\pm$3.0}}     & 73.1$\pm$16.3 & 7.5$\pm$3.3    \\
			\multicolumn{1}{l|}{DDFseg}   &  \textbf{66.9$\pm$11.0} & \multicolumn{1}{l|}{6.8$\pm$4.6}     & 79.1$\pm$6.7 & \multicolumn{1}{l|}{\textbf{6.6$\pm$3.9}}     & \textbf{83.5$\pm$16.0} & \multicolumn{1}{l|}{8.3$\pm$4.2}     & \textbf{76.5$\pm$13.8} & \textbf{7.3$\pm$4.3} \\ \hline
		\end{tabular}
		
	}%
	\label{ddfsegtab:2}
\end{table}

Table \ref{ddfsegtab:2} presents the comparison results on CT-MR cross modality segmentation.
One can see that U-Net (NoAdapt) performed poorly on both tasks because of the domain shift.
When MR images were taken as the target data with CT as the source domain, DDFseg achieved the best Dice and ASD values among all UDA methods on both tasks.
Particularly on MYO, when compared to SIFA, it obtained more than 4\% higher Dice score for MR segmentation and more than 10\% for CT segmentation.

\begin{table}[!t]
	\centering
	\caption{Comparison results for DDFSeg on LGE CMR segmentation. This table adopted from Pei et al. \cite{PEI2021102078}}
	\resizebox{\textwidth}{!}{ %
		\begin{tabular}{l|ll|ll|ll|ll}
			\hline
			\multirow{2}{*}{Method} & \multicolumn{2}{c|}{MYO}                             & \multicolumn{2}{c|}{RV}                              & \multicolumn{2}{c|}{LV}                              & \multicolumn{2}{c}{Mean}                            \\ \cline{2-9}
			& \multicolumn{1}{c}{Dice(\%)} & \multicolumn{1}{c|}{ASD(mm)} & \multicolumn{1}{c}{Dice(\%)} & \multicolumn{1}{c|}{ASD(mm)} & \multicolumn{1}{c}{Dice(\%)} & \multicolumn{1}{c|}{ASD(mm)} & \multicolumn{1}{c}{Dice(\%)} & \multicolumn{1}{c}{ASD(mm)} \\ \hline
			Unet(supervised) &  74.4$\pm$10.0 & 2.0$\pm$1.8  & 78.6$\pm$12.0     &  1.8$\pm$1.0   &  87.1$\pm$9.0   & 1.1$\pm$0.5       & 80.0$\pm$11.6  &1.7$\pm$1.3            \\\hline
			Unet(NoAdapt) & 29.6$\pm$19.5& 5.4$\pm$5.3& 48.1$\pm$20.2 & 3.5$\pm$1.9& 62.7$\pm$18.0 &3.4$\pm$1.8 &46.8$\pm$23.5&4.1$\pm$3.5   \\
			CycleSeg &57.1$\pm$14.9 &2.6$\pm$1.7& 75.7$\pm$15.3  &2.5$\pm$2.1 &82.2$\pm$9.3 &2.2$\pm$1.2 & 71.1$\pm$17.1 &2.4$\pm$1.7      \\
			SIFA & 68.1$\pm$15.0 & 2.2$\pm$1.9 &73.6$\pm$18.7 & 1.7$\pm$1.2 &83.5$\pm$13.0 &1.6$\pm$0.8 &75.1$\pm$16.9&1.8$\pm$1.4      \\
			CFDnet & 69.5$\pm$9.2 & 2.5$\pm$1.8 &77.6$\pm$8.8 & 1.9$\pm$1.4 &86.4$\pm$5.6 &1.9$\pm$0.9 &77.8$\pm$10.6&2.1$\pm$1.4     \\			
			DDFseg&  \textbf{75.0$\pm$7.3}& \textbf{1.4$\pm$1.3} & \textbf{84.5$\pm$7.0}&\textbf{1.3$\pm$0.8} &\textbf{88.6$\pm$5.0} & \textbf{1.4$\pm$0.9}  & \textbf{82.7$\pm$8.6} & \textbf{1.3$\pm$1.0} \\ \hline
		\end{tabular}
	}%
	\label{ddfsegtab:1}
\end{table}

For LGE CMR segmentation, we presents the comparison results in Table \ref{ddfsegtab:1}.
One can see that U-Net (NoAdapt) failed on this task, especially on myocardium.
This indicates the large domain shift between the two domains.
Among all the three UDA methods, DDFSeg achieved the best results, with 7.6\% higher average Dice score than SIFA.
Cycleseg and SIFA obtained comparable Dice scores on RV and LV, but the Dice scores were much lower than that of DDFSeg.
The reason could be that their performances were heavily dependent on the generated fake images.
However, they did not pay attention on the DSFs to enhance the translation process, while we introduced the zero-loss, which led to better image disentanglement and higher image quality of translated images.

\begin{table}[!t]
	\centering
	\caption{Performance comparisons for CFDNet on the CT-MR cross-modality cardiac segmentation task. This table adopted from Wu et al. \cite{Wu2020TMI}}
	\label{cfdnettable3}
		\resizebox{\textwidth}{25mm}{\begin{tabular}{|l|l|c|c|c|c|}
			\hline
			\multirow{2}{*}{Segmentation task} &\multirow{2}{*}{methods}&\multicolumn{2}{c|}{LV}&\multicolumn{2}{c|}{MYO}\\
			\cline{3-6}
			&&Dice(\%)&ASSD(mm)&Dice(\%)&ASSD(mm)\\
			\hline\hline
			Target: MR seg&		NoAdapt&44.4$\pm$13.9&19.1$\pm$6.52&24.4$\pm$9.11&17.3$\pm$2.70\\
			%\hline	
			Source: CT &		PnP-AdaNet&86.2$\pm$6.46&\textbf{2.74$\pm$1.04}&57.9$\pm$8.43&\ \ \textbf{2.46$\pm$0.661}\\
			%\hline
			&		AdvLearnNet&83.8$\pm$10.3&5.76$\pm$6.07&61.9$\pm$15.2&3.79$\pm$2.23\\	
			&		CORALnet&\textit{88.4$\pm$7.11}&3.02$\pm$2.45&\textit{67.7$\pm$10.8}&\textit{3.18$\pm$1.65}\\			
			%\hline			
			&		MMDnet&86.7$\pm$8.65&3.64$\pm$3.47&64.4$\pm$12.0&3.85$\pm$2.39\\
			%\hline
			&		CFDnet&\textbf{88.7$\pm$10.6}& \textit{2.99$\pm$2.79}&\textbf{67.9$\pm$8.62}&3.40$\pm$2.75\\
			\hline\hline

			Target: CT seg&	NoAdapt&30.3$\pm$27.7& N/A &0.140$\pm$0.130&N/A\\
			Source: MR &			PnP-AdaNet&\textit{78.3$\pm$18.4}&\textit{3.88$\pm$4.09}&\textit{62.8$\pm$8.24}&\textbf{3.09$\pm$1.59}\\
			%\hline
			&		AdvLearnNet&77.7$\pm$18.0&4.56$\pm$3.68&54.7$\pm$9.51&3.65$\pm$1.31\\	
			&		CORALnet&76.1$\pm$16.9&12.2$\pm$10.2&58.1$\pm$10.9&5.89$\pm$2.97\\			
			%\hline			
			&		MMDnet&77.7$\pm$18.2&5.62$\pm$4.86&57.1$\pm$12.0&4.13$\pm$1.70\\
			%\hline
			&		CFDnet&\textbf{81.9$\pm$18.2}& \textbf{3.64$\pm$3.94}&\textbf{62.9$\pm$10.9}&\textit{3.16$\pm$1.18}\\
			\hline
	\end{tabular}}
\end{table}

\begin{table*}[!t]
	\centering
	\caption{Performance comparison for CFDNet on LGE CMR images with bSSFP CMR as the source domain. This table adopted from Wu et al. \cite{Wu2020TMI}
	} \label{cfdnet bSSFP to lge table}
	\resizebox{\textwidth}{!}{
		\begin{tabular}{|l|c|c|c|c|c|c|}
			\hline
			\multirow{2}{*}{methods} & \multicolumn{3}{c|}{Dice (\%)} & \multicolumn{3}{c|}{ASSD (mm)}  \\
			\cline{2-7}
			&MYO  & LV & RV  & MYO  & LV & RV   \\
			\hline
			NoAdapt&14.5$\pm$20.12   &34.5$\pm$31.6  &31.1$\pm$26.3  &21.6$\pm$19.4  &11.3$\pm$13.1  &14.5$\pm$17.3 \\
			\hline	
			PnP-AdaNet &64.6$\pm$16.4  &78.4$\pm$16.2  &72.6$\pm$19.0 &4.64$\pm$6.41  &13.8$\pm$10.3  &5.30$\pm$5.33   \\
			\hline
			AdvLearnNet &65.5$\pm$13.7  &84.6$\pm$8.26  &\textit{75.2$\pm$16.5} &2.68$\pm$1.23  &3.70$\pm$2.33  &\textbf{4.08$\pm$2.65}   \\			
			\hline	
			CORALnet &\textit{68.0$\pm$10.4}  &\textit{85.2$\pm$6.41 } &73.8$\pm$11.7 &\textbf{2.30$\pm$0.831}  &3.43$\pm$1.66  &5.44$\pm$2.55   \\			
			\hline		
			MMDnet &67.0$\pm$9.83 &84.8$\pm$6.26  &72.3$\pm$11.4 &\textit{2.32$\pm$0.664}  &\textit{3.26$\pm$1.27}  &5.74$\pm$2.46   \\
			\hline
			CFDnet &\textbf{69.1$\pm$9.69}  &\textbf{86.4$\pm$5.62} &\textbf{76.0$\pm$10.9} &2.46$\pm$0.840 &\textbf{3.07$\pm$1.66}  &\textit{4.50$\pm$2.13}   \\
			\hline
	\end{tabular}}
\end{table*}

\begin{figure}[t]  \centering
	\includegraphics[width=1.0\textwidth]{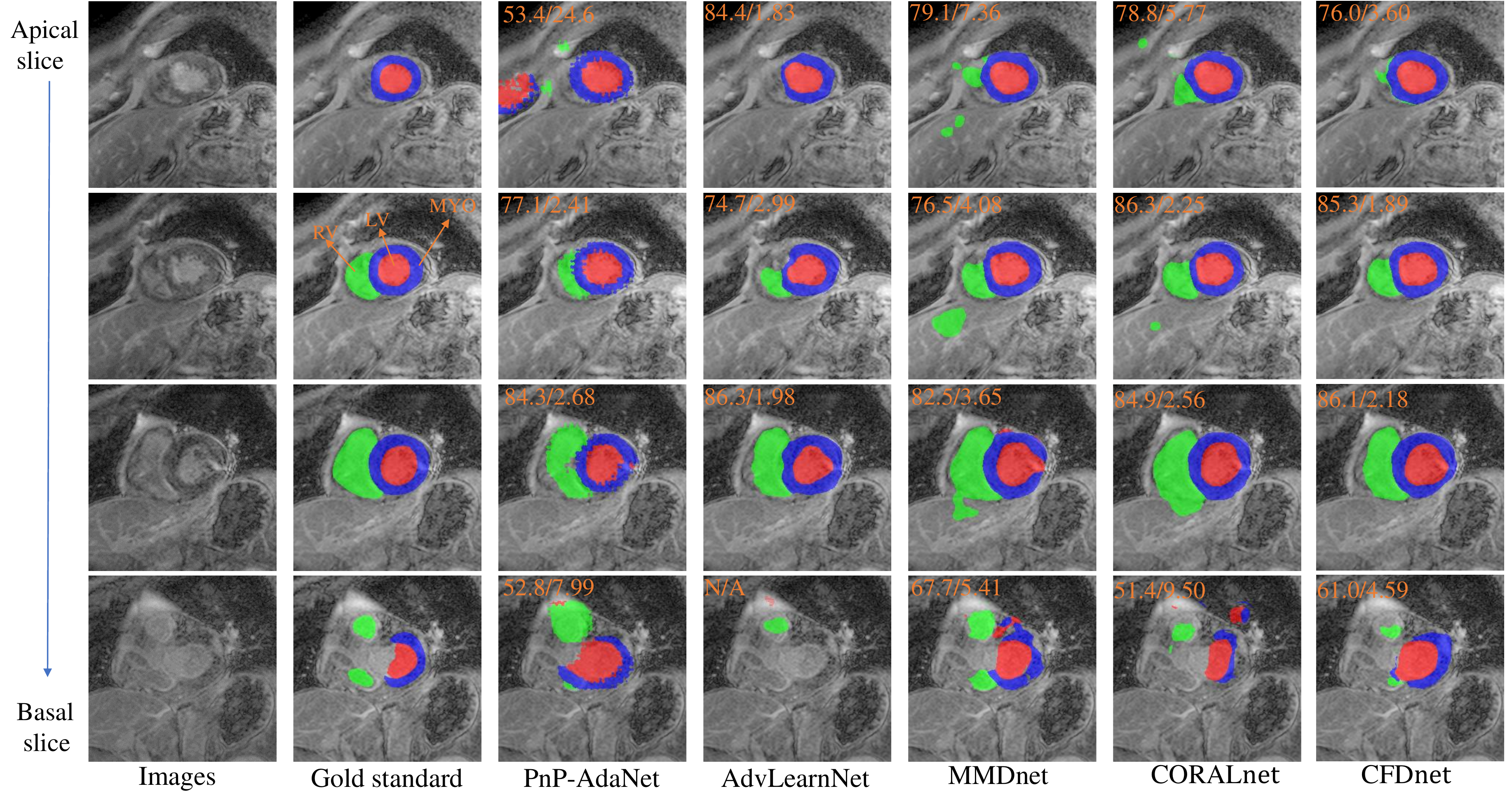}
	\caption{Visualization of 2D LGE CMR slices and segmentation results from comparison study for CFDNet. These are extracted from the test subject with median Dice score by CFDnet. Image adopted from Wu et al. \cite{Wu2020TMI}}
	\label{figcfdnet:LGE} \end{figure}	
\subsubsection{Comparison study for CFDNet}

We compared the the proposed CFDnet with five other methods: 
(1) NoAdapt, 
(2) PnP-AdaNet, 
(3) AdvLearnNet: We used the same network via adversarial training, 
(4) CORALnet: This method is the same as CFDNet  except using the distance between the second-order statistics (covariances) of the source and target features as the explicit metric, which was proposed by \cite{sun2016deep}, and 
(5) MMDnet: This method is the same as CFDNet except using MMD with Gaussian kernel function as the explicit metric.

Table \ref{cfdnettable3} presents the comparison results on CT-MR cross-modality cardiac segmentation.
When we took MR images as the target domain and CT as the source one, CFDNet achieved comparable results with PnP-AdaNet, though worse in ASSD values.
When tested on CT images with MR as the source domain, CFDnet obtained better in Dice scores, especially significantly better on LV ($p < 0.01$).
These results indicated that the proposed CF distance was effective for domain adaptation on segmentation tasks, and could achieve no worse prediction than the conventional adversarial training methods.

For LGE CMR segmentation, as shown in Table \ref{cfdnet bSSFP to lge table}, CFDNet obtained much better accuracies on all structures in all metrics compared to PnP-AdaNet and AdvLearnNet.
This results further demonstrated the effectiveness of explicit metric for domain adaptation.
Compared to other explicit metrics, CF distance obtained higher score on RV, and comparable results on LV and MYO.
This might be due to the substitution process for the computation of CF distance, which used the marginal distributions instead of joint ones.
This substitution leaded to a weaker constraint for feature extraction.
We further provided their visual comparison in Figure \ref{figcfdnet:LGE}.
One can see that CFDNet can achieve better prediction, while the shapes of the segmentation results were not satisfactory as expected.

\begin{table*}[t]
	\centering
		\caption{Performance comparison for VarDA on LGE CMR images. This table adopted from Wu et al. \cite{Wu2021TMI}}
		\label{bSSFP to lge table varda}
		\resizebox{\textwidth}{!}{
			\begin{tabular}{|l|c|c|c|c|c|c|}
				\hline
				\multirow{2}{*}{methods} & \multicolumn{3}{c|}{Dice (\%)} & \multicolumn{3}{c|}{ASSD (mm)}  \\
				\cline{2-7}
				&MYO  & LV & RV  & MYO  & LV & RV   \\
				\hline
				NoAdapt&14.50$\pm$20.18   &34.51$\pm$31.62  &31.10$\pm$26.30  &21.6$\pm$19.4  &11.3$\pm$13.1  &14.5$\pm$17.3 \\
				\hline
				PnP-AdaNet&64.64$\pm$16.41  &78.43$\pm$16.24  &72.66$\pm$19.04 &4.64$\pm$6.41  &13.8$\pm$10.3  &5.30$\pm$5.33   \\
				\hline
				CFDnet&69.1$\pm$9.69  &86.4$\pm$5.62 &76.0$\pm$10.9 &2.46$\pm$0.840 &3.07$\pm$1.66 &4.50$\pm$2.13\\
				\hline
				SIFA&70.66$\pm$9.689  &84.62$\pm$7.760  &~\textbf{83.99$\pm$6.821}  &~2.40$\pm$1.22  &~~2.68$\pm$1.14  &\textbf{2.05$\pm$1.19}   \\
				\hline
				VarDA&\textbf{73.03$\pm$8.316}  &\textbf{88.06$\pm$4.832}  &78.47$\pm$14.86   &~~\textbf{1.73$\pm$0.560}  &~\textbf{2.55$\pm$1.18}  &~3.51$\pm$2.24   \\
				\hline
		\end{tabular}}
\end{table*}

\begin{figure*}[!h]
	\centering
	\includegraphics[width=1.0\textwidth]{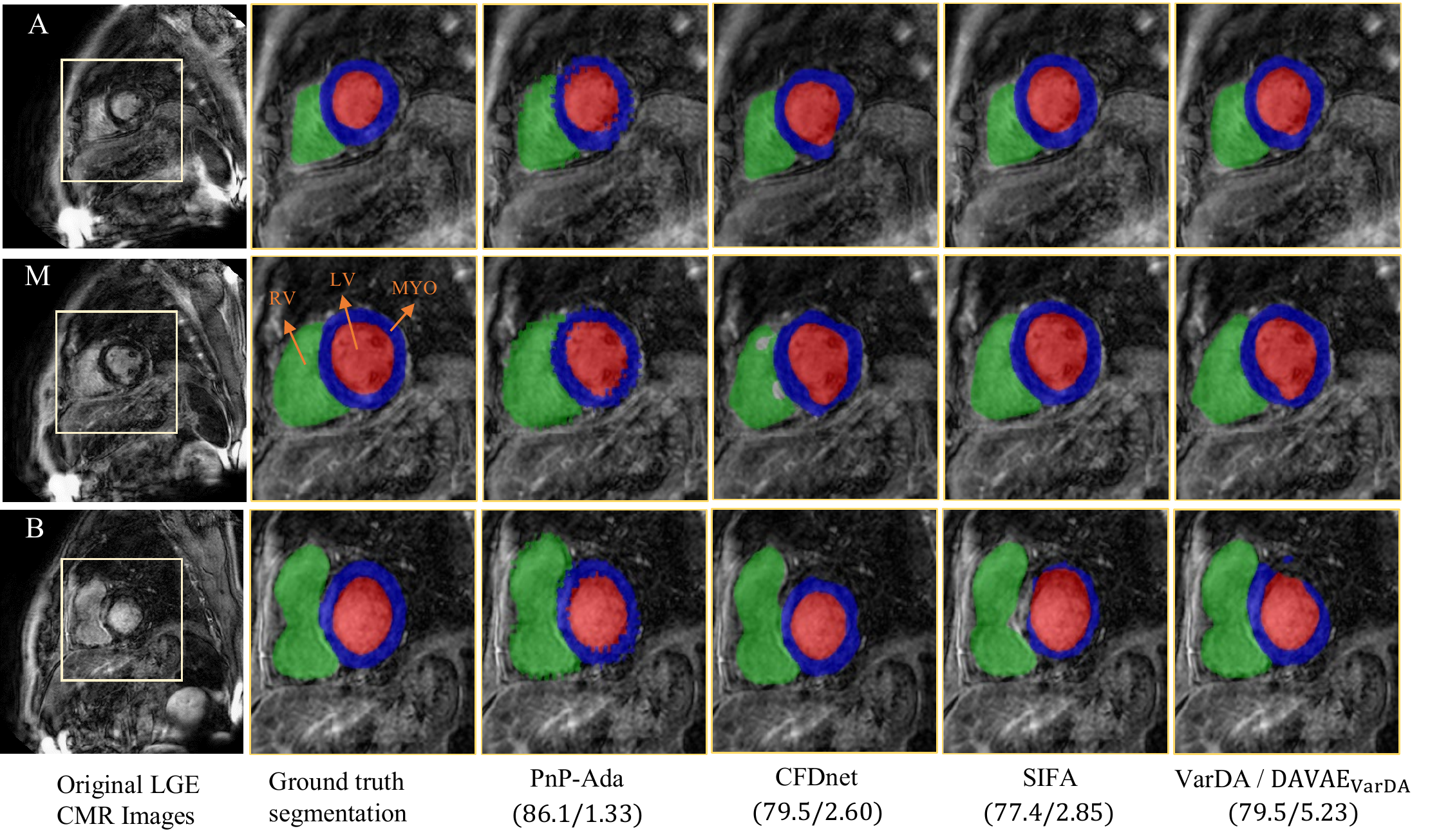}%\\[-2ex]
	\caption{
		Visualization of 2D LGE MR cardiac image segmentation results for comparison study for VarDA. 
		The cardiac structure of MYO, LV, RV are indicated in blue, red and green color, respectively. 
		Note that VarDA is the same as DAVAE$_{\mathrm{VarDA}}$. Image adopted from Wu et al. \cite{Wu2021TMI}
	}
	\label{figvarda:LGE}
\end{figure*}	

\subsubsection{Comparison study for VarDA}

We compared the proposed VarDA with three state-of-the-art methods, i.e., PnP-AdaNet (also denoted as PnP-Ada for short) \cite{dou2018pnp}, SIFA \cite{chen2019synergistic} and CFDNet.

Table \ref{bSSFP to lge table varda} presents the comparison results on LGE CMR segmentation.
The segmentation was done slice-by-slice in a 2D manner.
%({\bfseries visualization results display})
Figure \ref{figvarda:LGE} provides the visualization of the segmentation results of a subject, which was the median case of VarDA according to the average Dice score.
One can see that the proposed VarDA performed much better the CFDNet, which also used an explicit metric for domain adaptation.
The reason could be that it used variational approximation, which forced the correlations between the elements of the latent features to be weak, and thus leaded to a reasonable substitution of the metric calculation.
When compared to SIFA, VarDA obtained better accuracies on MYO and LV segmentation but worse on RV segmentation.

\subsection{Conclusion}

The three proposed domain adaptation frameworks investigated different aspects.
DDFSeg studied the technique of feature disentanglement by paying more attention on DSFs, and demonstrated that DSFs are useful for image translation and thus for domain adaptation.
CFDNet proposed a new metric for domain discrepancy, and was effective in cross-modality segmentation, with comparable results as the conventional adversarial training methods.
Based on CFDNet, we further proposed VarDA, which improved the effectiveness and performance of the explicit metric for domain adaptation.

\bibliographystyle{elsarticle-num}
%\normalem
\bibliography{chapter1}
%\begin{thebibliography}{00}
%
%\end{thebibliography}

\end{document}